# Pulse-Driven Self-Reconfigurable Meta-Antennas

Daiju Ushikoshi[1]*, Riku Higashiura[1]*, Kaito Tachi[1]*, Ashif Aminulloh Fathnan[1]*, Suhair Mahmood[1], Hiroki Takeshita[1], Haruki Homma[1], Muhammad Rizwan Akram[1], Stefano Vellucci[2], Jiyeon Lee[3], Alessandro Toscano[2], Filiberto Bilotti[2], Christos Christopoulos[4] and Hiroki Wakatsuchi[1,5]**

[1]Department of Engineering, Nagoya Institute of Technology, Gokiso-cho, Showa, Nagoya, Aichi, 466-8555, Japan.

[2]Department of Industrial, Electronic, Mechanical Engineering, ROMA TRE University, Rome, 00146, Italy.

[3]Electrical and Computer Engineering Department, University of California, San Diego, La Jolla, California 92093, USA.

[4]The George Green Institute of Electromagnetics Research, Department of Electrical and Electronic Engineering, The University of Nottingham, University Park, Nottingham, NG7 2RD, UK.

[5]Precursory Research for Embryonic Science and Technology (PRESTO), Japan Science and Technology Agency (JST), Kawaguchi, Saitama, 332-0012, Japan.

*These authors contributed equally to this work.
** Corresponding author. E-mail: wakatsuchi.hiroki@nitech.ac.jp.

((Abstract text))

Wireless communications and sensing have notably advanced thanks to the recent developments in both software and hardware. Although various modulation schemes have been proposed to efficiently use the limited frequency resources by exploiting several degrees of freedom, antenna performance is essentially governed by frequency only. Here, we present a new antenna design concept based on metasurfaces to manipulate antenna performances in response to the time width of electromagnetic pulses. We numerically and experimentally show that by using a proper set of spatially arranged metasurfaces loaded with lumped circuits, ordinary omnidirectional antennas can be reconfigured by the incident pulse width to exhibit directional characteristics varying over hundreds of milliseconds or billions of cycles, far beyond conventional performance. We demonstrate that the proposed concept can be applied for sensing, selective reception under simultaneous incidence and mutual communications as the first step to expand existing frequency resources based on pulse width.

1. Introduction

Wireless communication technologies have advanced at an accelerated rate, especially in recent years, due in part to the increasing demands for next-generation applications/services such as the Internet of Things (IoT), fifth-generation mobile communication systems (5G) and beyond 5G (B5G or 5G+), which manage a substantial number of communication devices within a single network.[1,2] In practice, communication devices are permitted to operate only in the frequency resources assigned.[3–5] To effectively utilize the assigned frequency resources, so far several modulation schemes have been proposed to multiplex the data over single channel by using frequency, time or code divisions to increase the throughput and hence accommodate more devices within the same wireless networks.[6,7] For instance, within a single wireless network using a given frequency resource, different signals can be sent and received without interference depending on the modulation scheme used as a frequency, time or code division multiple access system. In other words, modulation schemes introduce additional degrees of freedom to control signals, enabling simultaneous communications and a more efficient use of the limited frequency resources. From the hardware viewpoint, various antennas have been developed and explored in depth to satisfy certain requirements, including the physical dimensions, weight, materials and costs, while maintaining the expected antenna performance metrics, such as radiation pattern, bandwidth, gain and desired input impedance to be suitable for applications such as wireless sensing, wireless power transfer, security and bioapplications.[8] In general, these performance metrics vary depending on the frequency; conversely, antennas transmit/receive signals in the same manner at a fixed frequency. However, if we were able to vary such performance metrics at the antenna level itself by introducing an additional degree of freedom, each network could accommodate more communication devices without electromagnetic interference, making, thus, a significant advancement in wireless communication technology.

To vary antenna performance, reconfigurability is a typical key factor to consider. In fact, reconfigurable antennas are known to exhibit radiation characteristics that can be varied by electrical, mechanical (physical) or optical tuning mechanisms even at a constant frequency.[9–12] However, these mechanisms often rely on external energy resources (e.g., direct-current or DC power supplies) that require regular maintenance and thus are not suitable for complicated systems containing a large number of wireless communication devices, as expected, for instance, in advanced IoT sensor networks for industrial plants, healthcare services and smart houses and cities.[13,14] In addition, the quest for the zero-power devices required by the current energy/climate challenges is ever growing. Therefore, we propose that reconfigurable yet passive antennas will better satisfy the emerging demands for next-generation wireless systems.

Such a passive reconfigurability may be artificially produced by engineered subwavelength structures referred to as metamaterials[15,16] and metasurfaces.[17–19] These structures enable a straightforward implementation of a variety of electromagnetic features, including negative refractive indices[15,16,20] and extremely large impedance surfaces,[17] by suitably designing subwavelength unit cells. In fact, antennas designed with metasurfaces have shown extraordinary performance/features even beyond conventional physical limitations, such as miniaturization,[21] enhanced aperture efficiency,[22] ultrawideband radiation characteristics[23,24] and super-resolution.[25,26] Additionally, in wireless communications, metasurfaces serve as smart reflective/transmissive surfaces to effectively control the communication characteristics, including bit error rates,[27–30] or to convert an incoming wave into a different wavefront.[18,19,29,31-34] A metasurface-coating antenna has also been conceptually introduced to reduce the level of interference and provide additional physical layer security in wireless communications.[35]

These metasurfaces artificially produce tailored electromagnetic properties but static or invariable responses with linear media. To realize reconfigurability nonlinear metasurfaces have been introduced,[36-39] which give an additional state of impedance or electromagnetic properties at the same frequency even without external DC power supplies or in a passive manner.[40,41] For instance, metasurfaces embedded with schottky diodes were recently reported to be capable of discriminating between different waves by sensing the waveforms or, more specifically, the pulse widths.[42–47] The antenna design enabled by introducing such waveform-selective metasurfaces may allow an additional degree of freedom than can provide high throughput to receive/transmit/sense electromagnetic waves in a narrow frequency band as opposed to conventional antennas that require a wide frequency band. Potentially, the selectivity of pulse width can be exploited for sharing the same frequency resources to increase the number of wireless communication devices within a single network.

Several studies have validated that radiation characteristics vary in accordance with the pulse width.[45,48] However, those studies were limited to numerical simulations without any experimental validation. Additionally, none of the previous studies considered more than one incident wave source since overlapping pulses were recognized as a single combined pulse that could not be separated by conventional waveform-selective metasurfaces.[42–48] Moreover, in ordinary communication environments, multiple signals propagate at the same time, which, in practice, constrains waveform-selective performance.[6,7] For these reasons, both numerical and experimental validations are necessary for alternative passive reconfigurable antenna design to independently address multiple signals at the same frequency and at the same time.

Here, we present a new antenna paradigm based on waveform-selective metasurfaces that permit antenna characteristics to be suitably reconfigured for different waves without external energy resources. In particular, we demonstrate the realization of such advanced control by spatially constructing different types of waveform-selective metasurfaces as non-uniform waveform-selective metasurfaces around a single omnidirectional antenna. Due to this spatial configuration, our design is both numerically and experimentally proved to be capable of selectively receiving/transmitting pulsed surface waves and free-space waves at the same frequency and at the same time in a passive manner. To this end, first we present a design of a fundamental single-line system where one of the three types of waveform-selective metasurfaces is deployed as a straight line between two grounded monopole antennas to control surface wave propagation. These waveform-selective metasurfaces are based on periodic hexagonal conducting patches that transmit/absorb surface waves depending on the frequency and pulse width. These three types of waveform-selective metasurfaces are then integrated as a combined selective multi-line system to vary the direction of surface wave propagation over an entire 2D surface, which confirms that our systems have variable radiation characteristics at the same frequency. We also show that the proposed concept can be extended to design a selective system for free-space wave control by assembling a monopole antenna with waveform-selective metasurfaces based on slit (or slot) structures to selectively radiate signals in free space. Moreover, we present three specific applications to steer a main beam as a passive variable sensor, to receive a signal under simultaneous incidence and to build a mutual selective communication system without a frequency change or an external energy source. Thus, our study advances the design concept of antennas and wireless communication environments at the same frequency by using a new degree of freedom, the pulse width.

## 2. Theory

The proposed antenna concept is based on waveform-selective metasurfaces as explained below. Waveform selectivity refers to an electromagnetic response that is strongly coupled to the transient responses observed in fundamental electric circuits, more specifically, DC circuits.[42–45] Because of this coupling, the difference in between pulses can be sensed by waveform-selective metasurfaces, a property which can be used to vary wave propagation as well as antenna characteristics. The left panel of Fig. 1a shows two grounded monopole antennas linked by a metasurface line composed of nine unit cells, each of which is connected to a neighbouring cell by a diode bridge (Figure S1, Table S1 and Table S2 contain the specific design parameters). Consistent with other ordinary metasurfaces,[17] this structure strongly responds to an incoming surface wave at a resonant frequency, concentrating induced electric charges near the gaps between conducting patches. However, the rectification performed by the diode bridges generates an infinite set of frequency components, including a zero-frequency component, which dominates over any other frequencies. Therefore, transient circuit responses appear even at a constant frequency by deploying additional lumped components such as a resistor paired by a capacitor or an inductor.[42–44] Specifically, the capacitor-based (C-based) waveform-selective metasurface using the top circuit of Fig. 1b stores pulsed energy in capacitors at first and later dissipates it in the resistors. Importantly, this absorption performance decreases at steady state because the capacitors are fully charged. As a result, a long surface wave propagates over the C-based waveform-selective metasurface more strongly than a short pulse does. In contrast, the inductor-based (L-based) waveform-selective metasurface using the middle circuit of Fig. 1b exhibits poor absorptance during an initial time period due to the electromotive force of the inductors. Hence, a short pulse can be effectively transmitted. However, this force gradually disappears because of the zero frequency originated by the diode bridge rectification so that a long pulse is strongly absorbed by the L-based waveform-selective metasurface. Moreover, the two types of circuit configurations can be combined in parallel as a parallel-type (parallel) waveform-selective metasurface (see the bottom of Fig. 1b), which strongly absorbs both a short pulse and a continuous wave (CW). In this parallel waveform-selective metasurface, the absorption is high during an initial time period and gradually decreases since the capacitor is charged up. The absorption then increases again due to the decrease of the electromotive force of the inductors. Therefore, an intermediate pulse is weakly absorbed and thus transmitted over the parallel waveform-selective metasurface. Thus, these three types of waveform-selective metasurfaces (i.e., C-based, L-based and parallel waveform-selective metasurfaces) can be used to control incident waves in different manners depending on both frequency and pulse width. It is noted that the transient responses of these three circuit types are predictable by a DC circuit simulation as well as by using analytical calculation[49] (see Figure S2). The time constants of the exponentially varying voltages as well as their trends can be determined at will, hence facilitating the designs of the waveform-selective metasurfaces.

Surface wave control has been broadly reported and exploited in antenna applications, while most of them deal with frequency selectivity only.[50–54] In contrast, here we propose a waveform-selective metasurface to control surface waves such that eventually different signals from three antennas can be differentiated even though the signals arrive simultaneously at the same frequency. As the first step, the concept of our antennas can be simplified by the equivalent circuit drawn in Fig. 1c. In this figure, a transmitting antenna (Tx) is represented by an alternating current (AC) source, a switch, and input impedance $Z_0$, while the impedance of a receiver (Rx) is represented by $Z_1$. Additionally, $Z_R$ accounts for the rest of the energy, e.g., the energy travelling to free space. The energy received by the receiver, however, varies depending on the transmission line[55] between the two antennas, which is determined by the waveform-selective metasurface line. Note that, for the sake of simplicity, the equivalent circuit here does not include additional circuit components to describe other effects, including the coupling between antennas and the waveform-selective metasurface. Nonetheless, these additional effects are fully considered by electromagnetic simulations as will be shown in the later parts of this study.

The transmission between the two antennas can be more precisely calculated by using other methods, such as ABCD matrices and the Friis transmission equation.[8] However, one may intuitively understand from the simplified equivalent circuit of Fig. 1c that the amount of energy transmitted to the receiver depends not only on $Z_0$ and $Z_1$ but also on the transmission line in between, which is determined by the type of waveform-selective metasurface and varies in behaviour with the incident pulse width. Additionally, connecting more metasurface lines to the transmitter potentially influences the radiation direction from the transmitter. In this manner, we propose that omnidirectional antennas can be reconfigured to be time-varying directional antennas, as demonstrated below.

## 3. Results

### 3.1. Fundamental Single-Line System

First, we numerically designed and evaluated waveform-selective metasurfaces using a co-simulation method of ANSYS Electronics Desktop 2020R2 (see "Simulation Method" of the Methods Section and Figure S3), which were experimentally validated (see "Measurement Samples" and "Measurement Methods" of the Methods Section and Figure S4). In the absence of waveform-selective responses, the monopole antennas used in Fig. 1 were numerically adjusted to efficiently radiate signals at approximately 2.4 GHz, as shown in Fig. 1d. However, the transmittance varied depending on the incident pulse width, as explained above and numerically demonstrated in Fig. 1e (see Figure S5 for power dependence and Figure S6 for pulse width dependence). In particular, our waveform-selective metasurfaces were also designed to operate at approximately 2.4 GHz by properly choosing design parameters including the conducting dimensions, substrate thickness, etc. Importantly, the pulsed sine waves used in this study (simply called pulses below) had very narrow spectra compared to the bandwidths of both the antennas and the waveform-selective metasurfaces, as shown in Fig. 1d (cf. the bandwidth of the pulse in Fig. 1e). This ensured that our antennas did not rely on the difference in the frequency spectrum of pulsed sine waves to change the transmission characteristics. With a limited power level, these metasurfaces showed the same transmittance in the time domain, as their diodes had not yet been turned on (Figure S7).

However, by increasing the input power to 10 dBm, each type of waveform-selective metasurface maximized transmittance at a different time period even at a fixed oscillating frequency of 2.4 GHz because of their transient, waveform-selective absorbing mechanisms. For instance, as seen in the left panel of Fig. 1f, when the L-based, the parallel and the C-based waveform-selective meatsurfaces were used between the transmitter and the receiver, a transmittance peak appeared near 30 ns, 300 ns and 10 μs, respectively. In other words, most of the energy of a short pulse, an intermediate pulse and a long pulse can be effectively transmitted by the L-based, the parallel and the C-based waveform-selective metasurfaces, respectively. Such transient transmittance was experimentally observed but with a small frequency change to 2.42 GHz as shown in the right panel of Fig. 1f (see Fig. 1a for one of the measured samples and Figure S8 for the frequency characteristics). Compared to the simulation results (the left panel of Fig. 1f) where each transmittance peak was at least 10 dB larger than the transmittances of the other curves, in the measurement results (the right panel) the vertical gaps between the transmittance peaks and the second highest transmittances were reduced to smaller values. This was in part because the lumped circuit components were soldered by hand, which led to a difference in the amount of the solder used and thus resulted in minor frequency shifts. With small frequency adjustments less than 40 MHz, larger contrasts between the three transmittances were experimentally obtained in Figure S9. More details are seen in Figure S10 and Figure S11. Note that these waveform-selective mechanisms can be characterized by the values of the circuit elements, such as capacitance and inductance.[49] For instance, the C-based waveform-selective metasurface shown in the right panel of Fig. 1f was used again in Fig. 1g but with $C$ increased from 1 nF to 1 μF and 100 μF. As a result, the transient response appeared even over the order of several hundred milliseconds, which corresponds to billions of cycles (see Figure S12 for related simulation results). Note that this ultra-transient response was readily designed and experimentally validated by replacing discrete capacitor components of the C-based waveform-selective metasurfaces (i.e., replacing $C$ of Fig. 1b with a larger capacitance). The other types of waveform-selective metasurfaces can operate in a similar time period by using larger circuit component values. Theoretically, conventional antennas without waveform-selective metasurfaces may be able to achieve the same performance by using an extremely large quality factor.[56] Realistically, however, this mechanism becomes ineffective due to the presence of even a minor lossy component.

### 3.2. Combined Selective Multi-Line System

Next, we numerically and experimentally demonstrated that the propagation direction of the transient antenna design introduced in Fig. 1 can be readily extended by increasing the number of variable transmission lines, namely, the number of waveform-selective metasurface lines. Fig. 2a shows that a transmitting antenna expressed by an AC source, a switch and input impedance $Z_0$ is connected to three variable transmission lines $Z_C$, $Z_L$ and $Z_P$ representing C-based, L-based and parallel waveform-selective metasurface lines, respectively. $Z_1$, $Z_2$ and $Z_3$ denote the impedances of the three receivers. This concept is specifically realized by the schematic and the measurement sample of Fig. 2b, where the three lines used in Fig. 1f are connected to form a Y shape. In this design, the transmitter is positioned at the centre to transmit a signal to the three receivers at the terminals of the metasurface lines. Fig. 2c shows both numerically and experimentally that each receiver featured a maximum transient transmittance at a different time period since each waveform-selective metasurface line efficiently transmitted the signal during a different time slot.

As another planar scenario, hexagonal patches were fully deployed on a conducting ground plane to demonstrate that radiation characteristics can be changed over the entire 2D surface in Fig. 2d. This structure had a transmitting monopole and three types of waveform-selective metasurfaces, as shown in the Y-shaped structure in Fig. 2b. However, each waveform-selective metasurface occupied one-third of the area around the transmitter so that the surrounding waveform-selective metasurface was changed by every 120 degrees of reference angle to the transmitter. Under these circumstances, due to the presence of the different waveform-selective metasurfaces, the surface wave generated by the omnidirectional transmitter was expected to propagate over the ground plane unidirectionally, unlike conventional static metasurfaces.[17,53] As shown in Fig. 2e, the change in the radiation pattern of the transmitter was measured with the simplified method using three monopole receivers. In such transmittance measurements, an average transient transmittance peak first appeared between 0.01 and 0.1 μs in the L-based waveform-selective metasurface receiver area (Rx3), and then another such peak appeared between 0.1 and 1 μs in the parallel waveform-selective metasurface receiver area (Rx2). Finally, a larger transient transmittance was measured in the C-based waveform-selective metasurface receiver area (Rx1) at approximately 10 μs. These results indicate that most of the energy of a short pulse, an intermediate pulse and a long pulse can be effectively radiated to the directions of the L-based, the parallel and the C-based waveform-selective metasurfaces, respectively. As fundamentally demonstrated in Fig. 1 and Fig. 2, these results confirm that the use of non-uniform waveform-selective metasurfaces can provide an additional degree of freedom to control surface waves and vary radiation characteristics even at the same frequency. Note that our antenna designs demonstrated in Fig. 2 were spatially constructed with different waveform-selective metasurfaces. This means that our antennas used the spatial dimensions as an additional degree of freedom, which is later exploited to selectively distinguish different waves even under simultaneous incidence.

**3.3. System for Free-Space Wave Control**

To demonstrate the applicability to waves propagating in a wireless network or in free space, we also show that the proposed concept can be used to design a selective system for free-space wave control, as illustrated in Fig. 3a. In this figure, as seen in Fig. 2a, a transmitter is represented by an AC source, a switch and input impedance $Z_0$. In the case of free-space wave control, however, C-based, L-based and parallel waveform-selective metasurfaces are used as simple planar layers in parallel to the transmitter and, thus, represented by variable shunt impedances $Z_C$, $Z_L$ and $Z_P$ instead of transmission lines that correspond to free space. $Z_1$, $Z_2$ and $Z_3$, similarly, are the impedances of the three receivers, respectively. To specifically design such a system, a transmitting monopole antenna (18 mm tall) is surrounded by a set of waveform-selective metasurface panels in Fig. 3b. Note that, for simplicity, this scenario used a ground plane and the mirror image of the monopole to readily evaluate the radiation characteristics of a dipole antenna suspended in free space, unlike the surface wave cases in Fig. 1 and Fig. 2. The waveform-selective metasurfaces used in Fig. 3b were based on slit (or slot) structures[44,57] and, thus, designed to strongly transmit incoming signals at a resonant frequency (see the structure and frequency characteristics in Figure S13 and Figure S14, respectively, and the design parameters in Table S3 and Table S4). However, this resonant mechanism was either maintained or disrupted depending on the three types of waveform-selective metasurfaces used (two panels each, as shown in Fig. 3b). Note that even though the overall metasurface size in two panels was comparable to the wavelength (the dimension was equivalent to 0.9λ × 0.65λ) the single meta-atoms are subwavelength (~0.25λ). Thus, the metasurface effectively worked as a homogenized surface for the monopole antenna.[21, 22]

To observe the change in radiation characteristics in a simple manner, we used three monopole receivers placed in front of each waveform-selective metasurface (see Figure S15 and Table S5 and Table S6 for details and design parameters). Under this circumstance, the waveform-selective metasurface-based antenna steered the main lobe of the radiation by increasing the input power to 30 dBm at 3.85 GHz, as plotted in Fig. 3c (the frequency dependences are presented in Figure S16). Note that compared to the surface wave demonstrations seen in Fig. 1 and Fig. 2, this configuration used a higher frequency near 3.85 GHz instead of the frequency region near 2.4 GHz to reduce the dimensions of the measurement setup. At first, the signal was most efficiently radiated out of the panels containing the L-based waveform-selective metasurfaces, as obtained by Rx1 in Fig. 3b and shown by the black curve in Fig. 3c. Then, the L-based waveform-selective metasurface gradually reduced the transmittance, but the parallel waveform-selective metasurface started strongly transmitting the signal, where Rx2 measured the largest transmittance near 1 μs (see the red curve in Fig. 3c). This large transmittance, however, then decreased, while the C-based waveform-selective metasurface eventually maximized the transmittance, as shown by the blue curve in Fig. 3c.

Next, we compared the simulation results with measurement results. The simulated results in Fig. 3c were entirely consistent with the measurements of the sample shown in Fig. 3d, as plotted in Fig. 3e, except minor discrepancies in, for instance, the magnitude of transmittance and shift of time scales. Such discrepancies arose primarily due to the presence of some additional factors in the experimental case, including extra parasitic circuit components. Nonetheless, these measurements ensured the experimental feasibility of the proposed concept for free-space waves.

Furthermore, to experimentally visualize the radiation pattern over a 2D plane, we rotated the hexagonal metasurface prism of Fig. 3d every 10º whilst monitoring the transmittance from a monopole receiver in a far-field distance as seen in Fig. 3f. Consistent with the above simulation and measurement results, the radiation patterns plotted in Fig. 3g indicated that the maximum antenna directivity was achieved towards three different angles depending on the time period of the signal. The radiation pattern also clarified that the corner direction of each metasurface panel gave the highest transmittance as seen, for example, in 0.02 μs (the black lines in Fig. 3g), where the maximum transmittance was observed at 60º angle corresponding to the corner direction of the L-based waveform-selective metasurface panels. The side-lobe level in this time was 10 dB lower than the peak, indicating that receiving antennas located in other directions would receive low-intensity signals from the transmitter (see Figure S17 for the full time-varying far-field pattern). Besides the above 2D far-field profile, to get a clearer picture of interference paths, such as crosstalk, we have estimated the coupling from the transmitter to a receiver via the other two external antennas, and the result is shown in Figure S18. The calculated crosstalk was much lower than the power through the main path while in the measurement they differed by at least 20 dB.

As a reference result, Figure S19 shows simulation results where all the transmittances remained the same and constant when the input power level was not large enough to turn the diodes on. Additionally, Figure S20 represents the simulation results using only one of the three waveform-selective metasurfaces for all six panels. As shown in Figure S20, transient transmittance varied if the incident power was sufficiently large, but all the receivers received the same amount of energy, which indicates that the radiation characteristic was transient but omnidirectional. Note that as mentioned in Fig. 2, the antenna design of Fig. 3 was also spatially constructed with different waveform-selective metasurfaces. In other words, spatial dimensions were used as an additional degree of freedom, which plays an important role to separate different pulsed signals even at the same time as demonstrated in the following part of our study. Also, as mentioned in Fig. 2, the results of Fig. 3 indicate that most of the energy of a pulsed signal is effectively radiated to different directions depending on the pulse width (e.g., a short pulse to the direction of the L-based waveform-selective metasurface receiver).

**3.4. Passive Variable Sensor**

Now that we have so far designed and validated the antennas to vary radiation characteristics at the same frequency depending on the pulse width of a surface wave and a free-space wave, we use the proposed metasurface-based antennas in three different situations to explore how such ultra-transient directional antennas can be used for practical applications, even under the restriction of a single frequency band due to a limited frequency resource. In Fig. 3, the antenna was shown to steer the main lobe or transmit a signal to different receivers, which fits in existing wireless communication environments to vary the radiation characteristics and avoid electromagnetic interference. However, we propose an additional application using the reflected waveform of the transmitted signal to detect a scattering object as a passive yet variable sensor. As shown in Fig. 4a, a copper plate (51 mm tall and 70 mm wide) was placed in front of the antenna shown in Fig. 3. Note that we changed the location of the copper plate, which also influenced the reflectance in the time domain as the radiation characteristics of the antenna depend on the pulse width. Therefore, these results were compared to the reflectance without the copper plate to detect the difference in between and where the plate was positioned.

Under these circumstances, as shown in Figure S21a and more clarified in Fig. 4b, the transient reflectance experimentally increased during different time slots depending on the conductor position. Specifically, in contrast to the result obtained without the copper plate (Figure S21a), when the copper plate was located in front of the L-based waveform-selective metasurface panels, which efficiently radiated an electromagnetic wave during the initial time period, there was a larger transient reflectance until 0.2 μs. Similarly, the transient reflectance increased at approximately 1 μs and 10 μs when the plate was positioned in front of the parallel and C-based waveform-selective metasurface panels, respectively. These differences in reflectance appeared even if more than one copper plate were deployed, as demonstrated in Fig. 4c and Figure S21b. As seen in the black curve of Fig. 4c, for instance, a reflectance increased during an initial time period and near 1 μs, when two copper plates were positioned in front of the L-based and parallel waveform-selective metasurfaces, since they independently increased transmittance during these time slots. These results indicate that multiple objects can be sensed by using several types of waveform-selective metasurfaces or non-uniform waveform-selective metasurfaces. Moreover, the distances between the copper plates and these waveform-selective metasurfaces were changed in Fig. 4d and Figure S21c. As seen in Fig. 4d, the increase in reflectance was relatively limited during an initial time period when the distance between one of the copper plates and the L-based waveform-selective metasurface increased. In contrast, the gap between the two curves in Fig. 4d was relatively small near 1 μs, since the distance of another copper plate to the parallel waveform-selective metasurface was fixed. Thus, these results demonstrate that the distance to a scattering object can be detected from the difference in reflected waveforms. Additional results are seen in Figure S22. In this figure, reflectances are shown as a function of the distance between the monopole antenna and a copper plate.

In conventional methods to detect the direction of a scattering object, an antenna or a sensor (or a radar) usually changes its direction or posture as seen in parabolic antennas, adjusts the input phase as seen in phased arrays or sweeps the frequency component as a chirp signal.[8] Without any of these active changes, however, our waveform-selective metasurface-based antennas can potentially detect the location of scattering objects since the antennas keep varying their main lobes depending on pulse width, which is exploited as a new degree of freedom in wireless communications. Also, the proposed method only used a single transmitter unlike phased arrays composed of several radiating elements and phase shifters or adjustment components. Moreover, no duplexer was needed to perform detection even though only a single antenna was used. It is noted that our prototype here preferentially selected only three different angles. To increase the angular resolution, the metasurface can be arranged as a polygon with a higher number of sides that are associated with different waveform-selective functionalities. The metasurface antenna here detected objects and instantly differentiated their distances, while the detection reliability was limited by the noise floor as seen in Fig. 4d. Therefore, the performance can be improved by several factors including the number of sides, complexity in the circuit realization and the monopole antenna setups.

## 3.5. Selective Reception under Simultaneous Incidence

The second application is the selective reception of simultaneous incidences. From the viewpoint of efficient communication systems/environments, many signals may travel at the same time to increase the entire spatial data transfer rate. However, none of the waveform-selective metasurfaces that have been reported thus far is capable of distinguishing different pulsed waves if the signals arrive at the same time since the pulses are received as a single combined signal.[42–48] We addressed this issue in Fig. 5, where the central grounded monopole (effectively dipole) shown in Fig. 3 was placed near different types of waveform-selective metasurfaces that were spatially constructed to face different directions. This implies that our configuration exploited spatial dimensions as an additional degree of freedom (see Fig. 3a). Therefore, the proposed antenna design can be used for selectively receiving a pulsed signal from a particular incident angle.

Fig. 5a shows the simulation model tested using the same configuration as that in Fig. 3 but using the central antenna as a receiver and the three external antennas as transmitters, each of which generated a sine wave of 3.85 GHz (30 dBm) with a different phase offset, specifically, -120, 0 and +120 degrees for Tx1, Tx2 and Tx3, respectively. In this case, if the three signals have the same input power level, the total power received at the central receiver becomes ideally zero (or practically low) since the three incident signals cancel out each other, unless these signals are selectively filtered by waveform-selective metasurfaces. Therefore, the phase offset of the external transmitters clarify which signal dominated others even at the same frequency, as well as the direction of the signal source. Additionally, the distance between the central receiver and the external transmitters was reduced to ensure a sufficiently large input power for the measurements (see Figure S23 using the previous far-field distance).

First, when only one of the transmitters generated a signal, the simulation result was obtained, as plotted in the left panel of Fig. 5b. According to this result, transient transmittance was maximized during different time ranges, depending on which transmitter sent the signal. In addition, when more than one transmitter was activated, multiple transmittance peaks appeared, as plotted in the right panel of Fig. 5b. In particular, the three-source case turned out to follow the largest values among the three individual curves (see the left panel of Fig.

5b). To more clearly analyse the primary signal source, the voltage of the received signal was investigated in the time domain. According to the phases calculated in Fig. 5c (or the time of zero voltage), only one of the three signals was found to be relatively dominant across three time periods (i.e., approximately 0.05, 0.45 and 10 µs). This result indicates that by combining non-uniform waveform-selective metasurfaces with angular (or spatial) characteristics, different signals can be selectively received at the same frequency, even if the signals enter the proposed antenna at the same time. In addition, experimental validation is provided in Fig. 5d, Fig. 5e and Fig. 5f. In these measurements, basically we used the same conditions as the ones applied to the simulations in Fig. 5a, Fig. 5b and Fig. 5c except for the input power adjusted to 36 dBm (see the detailed measurement method in Figure S24 and "Measurement Methods" in the Methods Section). First, as seen in Fig. 5b, the measurement results seen in Fig. 5e showed similar transmittance peaks with minor shifts in the time domain due to several factors including the absence of solder and some additional parasitic circuit parameters that were not fully included in the simulations and affected the time constants of Fig. 5e. Despite these time-domain shifts, however, a single transmitted signal was found to dominate over the others, as shown in Fig. 5f. When all the three transmitters generated signals, the magnitude of the transmitted voltage slightly changed due to the interference between all the three signals. However, the phase remained almost the same as that of each dominant signal. These results experimentally confirm that the proposed antenna is capable of selectively receiving different pulses even under simultaneous incidences. Note that, as mentioned above, conventional design of waveform-selective metasurfaces or related antennas suffered from the fact that multiple pulsed signals arriving at the same time appeared as a single combined pulse that could not be distinguished by previous waveform selectivities. However, this drawback is overcome in the proposed antenna design concept by using spatial dimensions as an additional degree of freedom.

## 3.6. Mutually Selective Communication System

As the third application, we show that waveform selectivity can be exploited by multiple antennas to design a mutually pulse-width-selective communication system. In Fig. 3 and Fig. 5, only one antenna was permitted to selectively transmit/receive a pulsed signal in accordance with the pulse width. In contrast, the system proposed here is one step closer to a realistic wireless communication environment where ideally several antennas are expected to transmit and receive signals at the same time to increase the data transfer rate of the entire space. To realize such a system, the waveform-selective metasurface panels used only for a transmitter in Fig. 3 can be deployed for the three external antennas, since these external antennas are also given waveform-selective radiation characteristics, which means that mutually pulse-width-selective communications are established between any pair of antennas.

For simplicity, such a communication system was demonstrated using surface waves together with the three waveform-selective metasurface lines used in Fig. 2b. However, the three metasurface lines were combined to form a triangle shape instead of a Y shape in Fig. 6a. Each corner had a grounded monopole so that all the paths between the three antennas were connected by either a C-based, an L-based or a parallel waveform-selective metasurface to independently use different types of waveform selectivities without severely interfering with the other paths. Additionally, the antennas were programmed to generate a sine wave (10 dBm at 2.36 GHz) with a phase offset of -120, 0 or +120 degrees.

Under these circumstances, when only one of the three antennas generated a signal, the transmittances to the other two receivers were maximized in different time slots, as shown in Fig. 6b. This demonstrates selective transmission to the receiving antennas. Additionally, Fig. 6c shows that different signals were selectively received in different time ranges. In other words, the communication system shown in Fig. 6a exhibited not only the selective transmission capability (as presented in Fig. 2 and Fig. 3) but also the selective reception capability (as shown in Fig. 5), both of which were observed in every single antenna element of Fig. 6a.

The proposed system was experimentally validated using the measurement sample of Fig. 6d and the method introduced in Fig. 5 (see Figure S24). As a result, similar trends were obtained in Fig. 6e, where signals were selectively transmitted to receivers. In addition, two transmittance peaks appeared in each case of Fig. 6f even if two signals were simultaneously excited. Note that the locations of the two peaks were close to the peaks of the single source cases (i.e., compare the blue curves to the black and the red curves). These measurement results showed minor differences with the simulation results in Fig. 6b and Fig. 6c in terms of the magnitudes of the transmittances and the locations of the transmittance peaks. Nevertheless, the proposed mutually pulse-width-selective communication system was both numerically and experimentally validated.

## 4. Discussion

In this study, we demonstrated a proof-of-concept of pulse-driven self-reconfigurable antennas that contained several types of waveform-selective metasurfaces. Despite their simple configurations, the proposed antennas enabled successful variation of the antenna characteristics depending on the pulse width even at the same frequency. Specifically, Fig. 1 and Fig. 2 showed antennas generating surface waves that propagated in accordance with pulse width, while Fig. 3 demonstrated how free-space wave radiation changed at a constant frequency. However, in order to facilitate application of the proposed antennas, performance may need to be further optimized. For instance, Fig. 4 showed the application of a prototype antenna to sense a scattering object by varying the pulse width. This sensing capability was limited due to reflection from the waveform-selective metasurface panels since only one of the three types of panels permitted a signal to radiate out, while the other two types reflected the signal to the transmitting monopole, which appeared as a relatively strong reflection (see the results without copper plates in Figure S21). This performance can potentially be improved by introducing strong nonreciprocity into the waveform-selective metasurfaces.[58] In this way, the signal from the transmitting antenna would pass through the metasurfaces, but only the propagation from the outside to the antenna (only the wave reflected from the outside) would be selectively filtered so that the reflectance curve without the conductor plate in Figure S21 is significantly lowered to recognize objects that require much higher sensitivities. Additionally, our antennas relied on the diodes of commercial products. For this reason, the minimum power and dynamic range to achieve waveform selectivities or exploit pulse-driven reconfigurability were limited, as the diodes used had a turn-on voltage and a break-down voltage of finite values (approximately 0.3 V and 7.0 V, respectively). Between these voltages, diodes can be used to rectify incoming signals and achieve waveform-selective antenna performance. In particular, the turn-on voltage is important in wireless communications. In Fig. 3, for example, the incident power level was set to 30 dBm, which was much higher than ordinary communication signal levels (e.g., -70 dBm). Within the choice of commercial products, Schottky diodes as used in this work have a lower turn-on voltage than PIN diodes, but have higher equivalent resistance that potentially adds an ohmic loss leading to a reduced radiation efficiency. Enhanced performance can be achieved by using advanced semiconductor technologies, such as the latest microfabrication process, especially to lower the turn-on voltage of the diodes. Moreover, we note that compared to the simulation results, the measurement results showed a reduced transmittance or a reduced efficiency due to parasitic losses introduced during the fabrication process such as soldering of the surface mount devices (SMD). In Figure S25 we estimated the reduced transmittance by deliberately adding parasitic losses in the simulation and showed that an additional 100-$\Omega$ series resistance yielded a comparable transmittance profile to the measurement. To enhance the efficiency of the metasurface, a reliable fabrication process may be used to minimize the parasitic losses from SMD soldering. Despite the possibility of improving performance, the concept of the proposed antenna designs provides a new degree of freedom to control electromagnetic waves or wireless communication signals even at the same frequency which has not been shown previously. This was demonstrated here as selective transmitters in Fig. 1 to Fig. 3, a passive variable sensor in Fig. 4, a selective receiver in Fig. 5 and a mutual selective communication system in Fig. 6.

In the geometrical context, the hexagonal prism shape composed of the metasurface panels in Fig. 3 was meant to show the controllable radiation pattern depending on time, which is extensible into more than three angles. The metasurface prism can be arranged as a regular polygon with a reduced side number such as an equilateral triangle. However, the prism side length determined by the dimensions of metasurface panels should be carefully adjusted to maintain appropriate distance between the transmitting monopole and the metasurface panels, avoiding destructive phase interference (see Figure S26 for the analysis on triangular metasurface prisms). Additionally, a regular polygon with a higher number of sides such as a decagonal prism may increase the polygon radius too much, which exponentially increases the power requirement to achieve waveform-selective profiles. As for relations with the distance between the receivers and the directivity of the transmissive antennas, Friis transmission equation may be used to briefly approximate the power of the secondary reflections from external antennas. The power received by an antenna $P_R$ in decibel is $P_R = P_T + G_T + G_R + 20\log(\lambda/4\pi d)$. Here, $P_T$ is the transmitted power, $G_T$ and $G_R$ are the gains of the transmitter and the receiver, respectively, and $20\log(\lambda/4\pi d)$ is the distance-dependent power reduction where $\lambda$ is the operating wavelength and $d$ is the distance between the two antennas considered. In the case of the metasurface in Fig. 3b, the power received by Rx2 via Rx1 was -28.45 dBm according to the Friis formula, considering that the simulated gain of the monopole antennas was $G_T = G_R = 3.25$ dB and the transmitted power was $P_T = 0$ dBm. This calculated received power was approximately the same as the measured power as seen in Figure S18d which was around -28 dBm, confirming that there was only limited influence of the secondary reflections from external antennas via the metasurface panels.

From the viewpoint of efficiently utilizing frequency resources, the performance of our antenna design depends on the number of pulses that are selectively received/transmitted. As presented above (e.g., Fig. 1g and Figure S11), the transient responses basically follow exponential functions determined by time constants.[49] For this reason, only three different pulses (short, intermediate and long pulses) were unambiguously distinguished by our prototype antenna up to 10 μs as shown in Fig. 5. Communication signals can be more efficiently used by increasing the number of separated pulses, which is achieved by adding more metasurface filters. However, such improvement may be hampered by several fundamental restrictions including the slow mechanism to obtain transmittance variation that is based on the transient responses following exponential functions. Although this slow variation may still be acceptable in a small number of filtering such as shown in this work, it is likely to impose tight limitations when a higher number of filters is used. Therefore, a new mechanism is needed to go beyond this ordinary waveform-selective response that is not limited by exponentially time-varying voltages of RLC circuits. Nonetheless, we demonstrated that the proposed antenna design is experimentally realizable with quite long pulses even beyond many practical pulse width ranges, reaching billions of cycles or several hundred milliseconds (Fig. 1g). Performance can be further extended by replacing the lumped circuit components since their values relate to time constants that determine how the waveform selectivity or transient response is characterized (see past studies with respect to more advanced pulse controls[59] and equivalent circuit models for time constants[49]).

Additionally, for the efficient use of frequency resources, different modulation schemes can be combined to accommodate more communication devices in a single wireless network.[6,7] In recent years, orbital angular momentum (OAM) has been widely explored in which additional degrees of freedom can be achieved at the same frequency by exploiting orbital momenta of waves along with the established spin momenta, i.e., left/right hand circularly polarized waves.[33,60] Although the OAM is simple in concept and efficient multiplexing can be achieved, implementing the OAM technique in an antenna through the use of a metasurface usually requires a large design thickness that is proportional to the wavelength, which is acceptable in the optical range but not at lower frequencies. Additionally, the challenging design of receiver antennas to detect OAM waves in the far-field makes it nearly impossible to utilize for long-distance communications. In contrast, our waveform-selective metasurfaces can be designed with a subwavelength thickness and be installed directly to an antenna anywhere in a wireless network environment.[45] Nevertheless, the concept of waveform selectivity can be integrated with the OAM as well as with other modulation schemes[6,7] to accommodate an increasing number of wireless communication devices in a single wireless network and to more effectively share the same frequency resource. With such a potential, our study is expected to advance the design concept of antennas and wireless communication environments at the same frequency and become the first step to expand existing frequency resources by using a new additional degree of freedom, the pulse width.

## 5. Conclusion

In conclusion, we have presented a design concept of metasurface-based antennas and their applications to reconfigure antenna characteristics at the same frequency without external energy resources, namely, in a passive manner. The proposed antennas exhibited ultra-transient, pulse-width-selective characteristics that enabled the realization of varied scattering parameters at a constant frequency over several hundred milliseconds. This capability is not practically achievable by conventional antenna design methods but was made possible by the utilization of waveform-selective metasurfaces. In particular, the use of spatially constructed non-uniform waveform-selective metasurfaces led to selective control of the radiation characteristics of ordinary omnidirectional antennas, which were validated both numerically and experimentally. Finally, these antennas were used in three applications, specifically to steer a main beam as a passive yet variable sensor, to receive a signal under simultaneous incidence and to build a mutually selective communication system without a frequency change or an external energy source. Unlike previous studies which were limited to simulations only and a single signal source, our study both numerically and experimentally verified selective transmission and reception even under simultaneous incidence as often seen in realistic communication environments. Our study is expected to provide a new degree of freedom to design wireless communication systems and to contribute to the efficient use of limited frequency resources in the next generation of communication environments.

**6. Methods**

*Simulation Method*: We designed and evaluated antennas based on waveform-selective metasurfaces by using a co-simulation method[42–44] that integrated an electromagnetic (EM) solver with a circuit simulator within ANSYS Electronics Desktop 2020 R2. In this method, first, the EM models were simulated in the EM solver without lumped circuit components that were replaced by lumped ports (Figure S3). The scattering parameters obtained in the EM solver were then used in the circuit simulator along with the lumped components connected to the lumped ports. This method is equivalent to directly connecting the lumped components to the EM models in the EM solver, although it significantly improves the simulation efficiency and contributes to optimizing the antenna models. We note that diodes were properly simulated using the corresponding SPICE model where nonlinearity was included. Unlike other simulation methods based on time-domain analysis such as the Finite Difference Time Domain (FDTD) method or the Transmission Line Modelling (TLM) method,[61] the co-simulation method performs transient analysis within the circuit level, which does not normally produce field distributions but more efficiently provides voltages and currents at discrete circuit components.

*Measurement Samples*: Our waveform-selective metasurfaces were fabricated using Rogers3010 substrates (1.27 mm thick) and Broadcom HSMS-286X series product diodes. The design parameter details are given in Table S1, Table S2, Table S3, Table S4, Table S5 and Table S6. The conducting patterns were covered by coating layers to prevent the influence of oxidation and to facilitate the soldering process. Additionally, the front surface of the measurement sample used in Fig. 2d was mostly covered by a resist that helped limit the solder area.

*Measurement Methods*: As shown in Figure S4, input signals were generated by a signal generator (Anritsu MG3692C) or an arbitrary waveform generator (AWG) (Keysight Technologies M8195A). In particular, the AWG was used to generate pulses longer than 10 μs, as shown in Fig. 1g. The duty cycles of the pulsed signals were set to 0.1 % or less to ensure that the pulse periods were relatively long enough, compared to the pulse width. This was because the electric potential within diode bridges needed to be restored to zero voltage before a next pulse came in. If the power levels were too small, an amplifier (Ophir 5193RF) was used to ensure that the incident power levels were sufficient to realize waveform-selective responses. The reflection to the amplifier was reduced by an isolator (Pasternack PE83IR1004). For the measurements of Fig. 5 and Fig. 6, a divider (Clear Microwave D38004), phase shifters (Pasternack PE8245) and a coupler (ET Industries C-058-30) were additionally used (for instance, see Figure S23 for the setup of Fig. 5). The incident, reflected and transmitted powers/energies were measured using an oscilloscope (Keysight Technologies DSOX6002A or Teledyne LeCroy WaveRunner9404 M). The frequency characteristics of the proposed antennas were measured by using a vector network analyser (VNA) (Keysight Technologies N5249A).

**Data availability**

The data that support the findings of this study are available from the corresponding author upon reasonable request.

**Code availability**

The codes that are used to generate results in the paper are available from the corresponding author upon reasonable request.

**References**
[1] Dang, S., Amin, O., Shihada, B. & Alouini, M.-S. What should 6G be? *Nat. Electron.* **3**, 20-29 (2020).

[2] Al-Fuqaha, A. et al. Internet of things: A survey on enabling technologies, protocols, and applications. *IEEE Commun. Surv. Tutorials* **17**, 2347-2376 (2015).

 [3] *THE EUROPEAN TABLE OF FREQUENCY ALLOCATIONS AND APPLICATIONS IN THE FREQUENCY RANGE 8.3 KHz to 3000 GHz (ECA TABLE) Approved November 2020*, 2020.

[4] *FCC ONLINE TABLE OF FREQUENCY ALLOCATIONS47 C.F.R. § 2.106 Revised on February 1, 2021*, 2021.

[5] *MIC Frequency Assignment Plan, September 2021*, 2021.

[6] Goldsmith, A. *Wireless Communications* (Cambridge Univ. Press, 2005).

[7] Hanzo, L., Ng, S. X., Webb, W. & Keller, T. *Quadrature Amplitude Modulation: From Basics to Adaptive Trellis-Coded, Turbo-Equalised and Space-Time Coded OFDM, CDMA and MC-CDMA Systems* (IEEE Press-John Wiley, 2004).

[8] Balanis, C. A. *Antenna Theory: Analysis and Design* (John Wiley & Sons, 2016).

[9] Hum, S. V. & Perruisseau-Carrier, J. Reconfigurable reflectarrays and array lenses for dynamic antenna beam control: A review. *IEEE Trans. Antennas Propag.* **62**, 183-198 (2014).

[10] Huang, C. et al. Reconfigurable intelligent surfaces for energy efficiency in wireless communication. *IEEE Trans. Wireless Commun.* **18**, 4157-4170 (2019).

[11] Christodoulou, C. G., Tawk, Y., Lane, S. A & Erwin, S. R. Reconfigurable antennas for wireless and space applications. *Proc. IEEE* **100**, 2250-2261 (2012).

[12] Wu, C. et al. A phased array based on large-area electronics that operates at gigahertz frequency. *Nat. Electron.* **4**, 757-766, (2021).

[13] Stankovic, J. A. Research directions for the internet of things. *IEEE Internet Things J.* **1**, 3-9 (2014).

[14] Zanella, A. et al. Internet of things for smart cities. *IEEE Internet Things J.* **1**, 22-32 (2014).

[15] Smith, D. R. et al. Composite medium with simultaneously negative permeability and permittivity. *Phys. Rev. Lett.* **84**, 4184 (2000).

[16] Shelby, R. A., Smith, D. R. & Schultz, S. Experimental verification of a negative index of refraction. *Science* **292**, 77-79 (2001).

[17] Sievenpiper, D. et al. High-impedance electromagnetic surfaces with a forbidden frequency band. *IEEE Trans. Microw. Theory Tech.* **47**, 2059-2074 (1999).

[18] Yu, N. et al. Light propagation with phase discontinuities: generalized laws of reflection and refraction. *Science* **334**, 333-337 (2011).

[19] Yu, N. & Capasso, F. Flat optics with designer metasurfaces. *Nat. Mater.* **13**, 139-150 (2014).

[20] Grbic, A. & Eleftheriades G.V. Overcoming the diffraction limit with a planar left-handed transmission-line lens. *Phys. Rev. Lett.* **92**, 117403 (2004).

[21] Ziolkowski, R. W., Jin, P. & Lin, C.-C. Metamaterial-inspired engineering of antennas. *Proc. IEEE* **99**, 1720, (2011).

[22] Binion, J. D. et al. A metamaterial-enabled design enhancing decades-old short backfire antenna technology for space applications. *Nat. Commun.* **10**, 1-7 (2019).

[23] Lier, E. et al. An octave-bandwidth negligible-loss radiofrequency metamaterial. *Nat. Mater.* **10**, 216-222 (2011).

[24] Sussman-Fort, S. E. & Rudish, R. M. Non-Foster impedance matching of electrically-small antennas. *IEEE Trans. Antennas Propag.* **57**, 2230-2241 (2009).

[25] Lerosey, G., de Rosny, J., Tourin, A. & Fink, M. Focusing beyond the diffraction limit with far-field time reversal. *Science* **315**, 1120-1122 (2007).

[26] Grbic, A. Jiang, L. & Merlin, R. Near-field plates: Subdiffraction focusing with patterned surfaces. *Science* **320**, 511-513 (2008).

[27] Wu, Q. & Zhang, R. Towards smart and reconfigurable environment: Intelligent reflecting surface aided wireless network. *IEEE Commun. Mag.* **58**, 106-112 (2019).

[28] Dunna, M., Zhang, C., Sievenpiper, D. & Bharadia, D. ScatterMIMO: Enabling virtual MIMO with smart surfaces. *Proc. 26th Annu. Int. Conf. Mob. Comput. Netw.* 1-14 (2020).

[29] Chen, L. et al. Pushing the physical limits of iot devices with programmable metasurfaces. *18th USENIX Symp. Networked Sys. Des. Implement.* 425–438 (2021).

[30] Zhang, L. et al. A wireless communication scheme based on space-and frequency-division multiplexing using digital metasurfaces. *Nat. Electron.* **4**, 218-227 (2021).

[31] Pfeiffer, C. & Grbic, A. Metamaterial Huygens' surfaces: tailoring wave fronts with reflectionless sheets. *Phys. Rev. Lett.* **110**, 197401 (2013).

[32] Shen, Y. et al. Optical vortices 30 years on: OAM manipulation from topological charge to multiple singularities. *Light Sci. Appl.* **8**, 1-29 (2019).

[33] Ren, H. et al. Complex-amplitude metasurface-based orbital angular momentum holography in momentum space. *Nat. Nanotechnol.* **15**, 948-955 (2020).

[34] Olk A. E., Macchi, P. E., & Powell, D. A. High-efficiency refracting millimeter-wave metasurfaces. *IEEE Trans. Antennas Propag.* **68**,5453-5462 (2020).

[35] Tsiftsis, T. A., Valagiannopoulos, C., Liu, H., Boulogeorgos, A. A. A., & Miridakis, N. I. Metasurface-coated devices: A new paradigm for energy-efficient and secure 6G communications. *IEEE Veh. Technol. Mag.* **17**, 27-36 (2022).

[36] Lapine, M., Shadrivov, I. V. & Kivshar, Y. S. Colloquium: Nonlinear metamaterials. *Rev. Mod. Phys.* **86**, 1093 (2014).

[37] Zharov, A. A., Shadrivov, I. V. & Kivshar, Y. S. Nonlinear properties of left-handed metamaterials. *Phys. Rev. Lett.* **91**, 037401 (2003).

[38] Shadrivov, I. V., Morrison, S. K. & Kivshar, Y. S. Tunable split-ring resonators for nonlinear negative-index metamaterials. *Opt. Express* **14**, 9344-9349 (2006).

[39] Renzo, M. D. et al. Smart radio environments empowered by reconfigurable intelligent surfaces: How it works, state of research, and the road ahead. *IEEE J. Sel. Areas Commun.* **38**, 2450-2525 (2020).

[40] Barbuto, M. et al. Metasurfaces 3.0: A new paradigm for enabling smart electromagnetic environments. *IEEE Trans. Antennas Propag.* (2021).

[41] Barbuto, M. et al. Intelligence enabled by 2d metastructures in antennas and wireless propagation systems. *IEEE Open J. Antennas Propag.* **3**, 135 (2022).

[42] Wakatsuchi, H., Kim, S., Rushton, J. J. & Sievenpiper, D. F. Waveform-dependent absorbing metasurfaces. *Phys. Rev. Lett.* **111**, 245501 (2013).

[43] Wakatsuchi, H. et al. Waveform selectivity at the same frequency. *Sci. Rep.* **5**, 1-6 (2015).

[44] Wakatsuchi, H., Long, J. & Sievenpiper, D. F. Waveform selective surfaces. *Adv. Funct. Mater.* **29**, 1806386 (2019).

[45] Vellucci, S. et al. Waveform-selective mantle cloaks for intelligent antennas. *IEEE Trans. Antennas Propag.* **68**, 1717 (2020).

[46] Imani, M. F. & Smith, D. R. Temporal microwave ghost imaging using a reconfigurable disordered cavity. *Appl. Phys. Lett.* **116**, 054102 (2020).

[47] Eleftheriades, G. V. Protecting the weak from the strong. *Nature* **505**, 490-491 (2014).

[48] Barbuto, M. et al., Waveguide components and aperture antennas with frequency-and time-domain selectivity properties. *IEEE Trans. Antennas Propag.* **68**, 7196-7201 (2020).

[49] Asano, K., Nakasha, T. & Wakatsuchi, H. Simplified equivalent circuit approach for designing time-domain responses of waveform-selective metasurfaces. *Appl. Phys. Lett.* **116**, 171603 (2020).

[50] Fong, B. H. et al. Scalar and tensor holographic artificial impedance surfaces. *IEEE Trans. Antennas Propag.* **58**, 3212-3221 (2010).

[51] Quarfoth, R. & Sievenpiper, D. Artificial tensor impedance surface waveguides. *IEEE Trans. Antennas Propag.* **61**, 3597-3606 (2013).

[52] Lee, J. & Sievenpiper, D. F. Method for extracting the effective tensor surface impedance function from nonuniform, anisotropic, conductive patterns. *IEEE Trans. Antennas Propag.* **67**, 3171-3177 (2019).

[53] Bosiljevac, M. Non-uniform metasurface Luneburg lens antenna design. *IEEE Trans. Antennas Propag.* **60**, 4065-4073 (2012).

[54] Minatti, G. et al. Synthesis of modulated-metasurface antennas with amplitude, phase, and polarization control. *IEEE Trans. Antennas Propag.* **64**, 3907-3919 (2016).

[55] Christopoulos, C. *The Transmission-Line Modeling Method* (IEEE Press, New Jersey, 1995).

[56] Nakasha, T., Phang, S. & Wakatsuchi, H. Pseudo-waveform-selective metasurfaces and their limited performance. *Adv. Theory Simul.* **4**, 2000187 (2021).

[57] Munk, B. A. *Frequency Selective Surfaces: Theory and Design* (John Wiley & Sons, 2005).

[58] Caloz, C. et al. Electromagnetic nonreciprocity. *Phys. Rev. Appl.* **10**, 047001 (2018).

[59] Wakatsuchi, H. Time-domain filtering of metasurfaces. *Sci. Rep.* **5**, 16737 (2015).

[60] Lloyd, S. M., Babiker, M., Thirunavukkarasu, G. & Yuan, J. Electron vortices: Beams with orbital angular momentum. *Rev. Mod. Phys.* **89**, 035004 (2017).

[61] Wakatsuchi, H., Anzai, D., & Smartt C. Visualization of field distributions of waveform-selective metasurface. *IEEE Antennas Wirel. Propag. Lett.* **15**, 690-693 (2015).

**Acknowledgements**

This work was supported in part by the Japanese Ministry of Internal Affairs and Communications (MIC) under the Strategic Information and Communications R&D Promotion Program (SCOPE) No. 192106007, the Japan Science and Technology Agency (JST) under the Precursory Research for Embryonic Science and Technology (PRESTO) No. JPMJPR193A, and Japan Society for the Promotion of Science (JSPS) KAKENHI No. 17KK0114 and No. 21H01324. FB, AT and SV acknowledge the support of the research contract MANTLES funded by the Italian Ministry of University and Research, PRIN 2017 No. 2017BHFZKH.

**Author contributions**

H.W. conceived of the idea and designed the project. D.U., R.H. and K.T. primarily performed simulations and measurements under the support of A.A.F., S.M., H.T., H.H., M.R.A., S.V., J.L. and H.W. All authors contributed to analyzing the results and editing the paper.

**Competing interests**

The authors declare no competing interests.

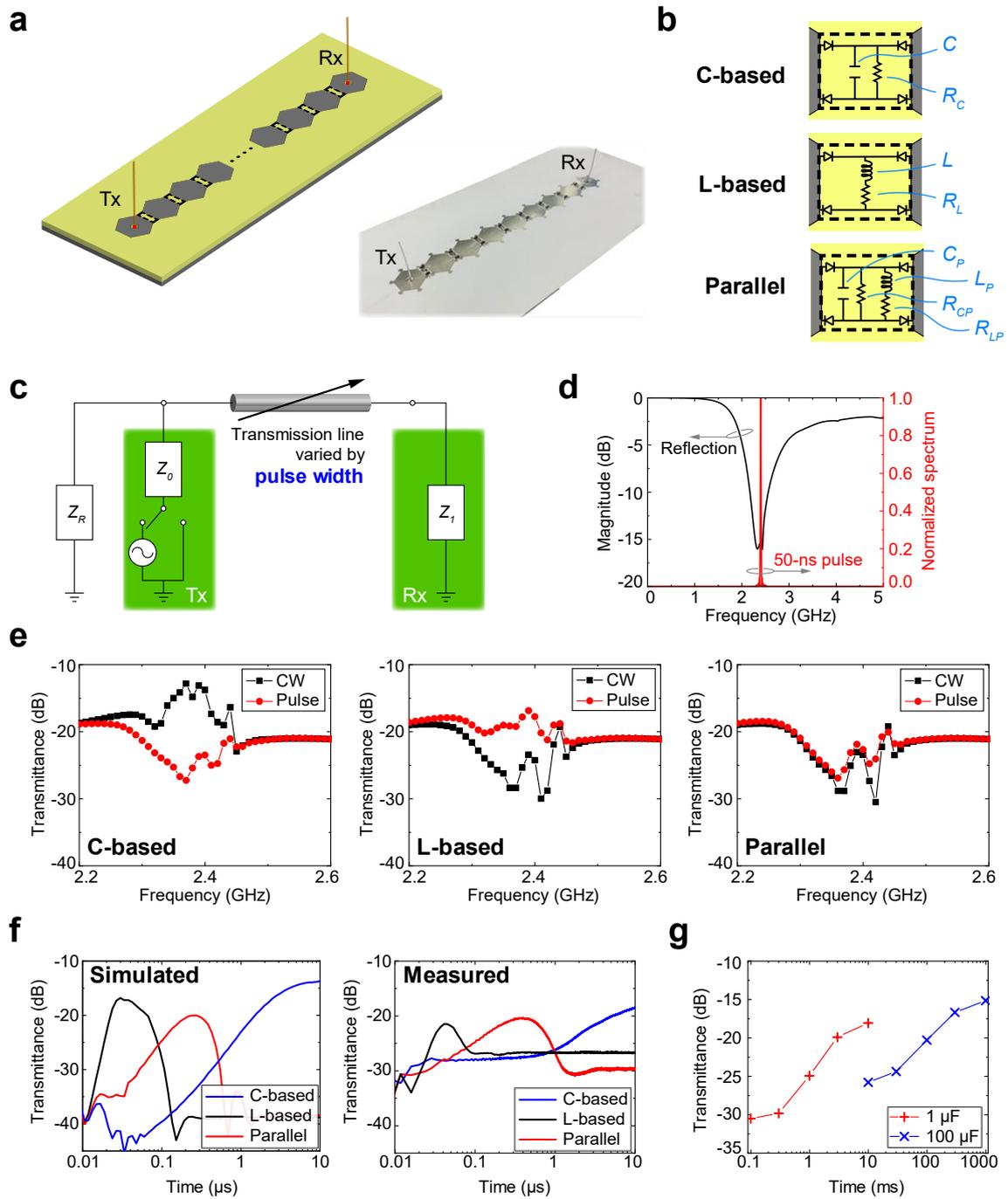

Fig. 1: Fundamental Single-Line System. (a) Simulation model and measurement sample. Ordinary grounded monopole transmitters (Tx) and receivers (Rx) were connected by waveform-selective metasurface lines to vary surface wave propagation and exhibit ultra-transient responses at the same frequency. (b) Circuit configurations for C-based, L-based and parallel waveform-selective metasurfaces. Design parameters are seen in Figure S1, Table S1 and Table S2. (c) Simplified equivalent circuit concept representing the two monopole antennas (Tx and Rx) connected by the waveform-selective metasurface line. (d) Simulated reflectance of the monopoles and normalized spectrum of a 50-ns sine wave pulse of 2.42 GHz. (e) Simulated transmittances between the antennas connected by (left) C-based, (centre) L-based, and (right) parallel-type waveform-selective metasurface lines as a function of frequency with a 10 dBm input power. In the legend, CW represents continuous wave as sufficiently long pulse, while Pulse indicates 50-ns-long short pulse (see Figure S6 for a comparison to an intermediate pulse of 300 ns). (f) The corresponding transient transmittances at 2.4 GHz in the (left) simulation and (right) measurement (also see Figure S6 for simulated transmittances with variation of pulse-width). (g) Measured results using larger capacitances for the C-based waveform-selective metasurface line. $C$ was increased from 1 nF to 1 μF and 100 μF.

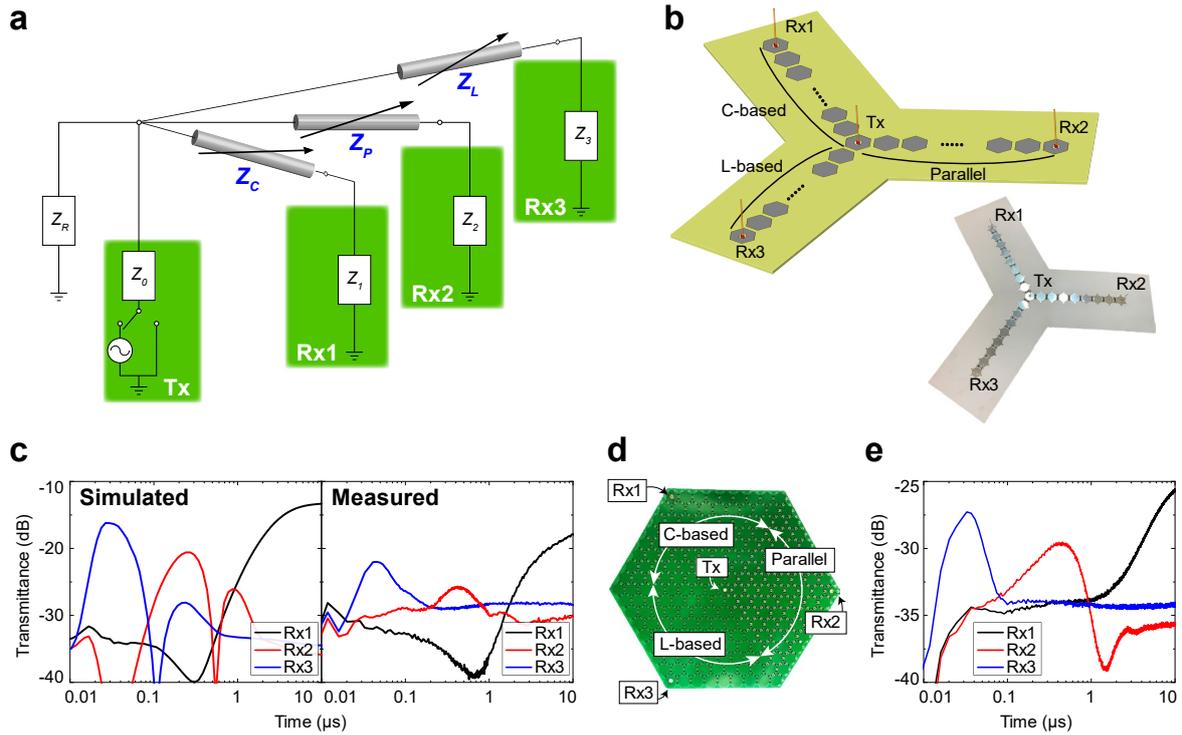

Fig. 2: Combined selective multi-line systems and associated performance. (a) Extended equivalent circuit concept using additional waveform-selective metasurface lines to selectively transmit signals. Three variable transmission lines $Z_C$, $Z_L$, and $Z_P$ represented C-based, L-based and parallel waveform-selective metasurface lines, while $Z_1$, $Z_2$, and $Z_3$ were impedances of three receivers Rx1, Rx2 and Rx3. (b) Monopole transmitter connected by all three types of waveform-selective metasurface lines. (c) (left) Simulated and (right) measured transmittances to Rx1, Rx2 and Rx3. The frequency was set to 2.36 GHz and 2.45 GHz in the simulation and measurement, respectively, while the input power was fixed at 15 dBm in both cases. (d) Image of a prototype with three waveform-selective metasurfaces fully extended on a 2D surface to vary radiation characteristics over the entire surface. (e) Corresponding measured transmittances at 2.24 GHz with 23 dBm. Note that the 2D metasurface case had a lower operating frequency due to a higher number of electronic circuits soldered to hexagonal patch sides, which introduced more parasitic elements.

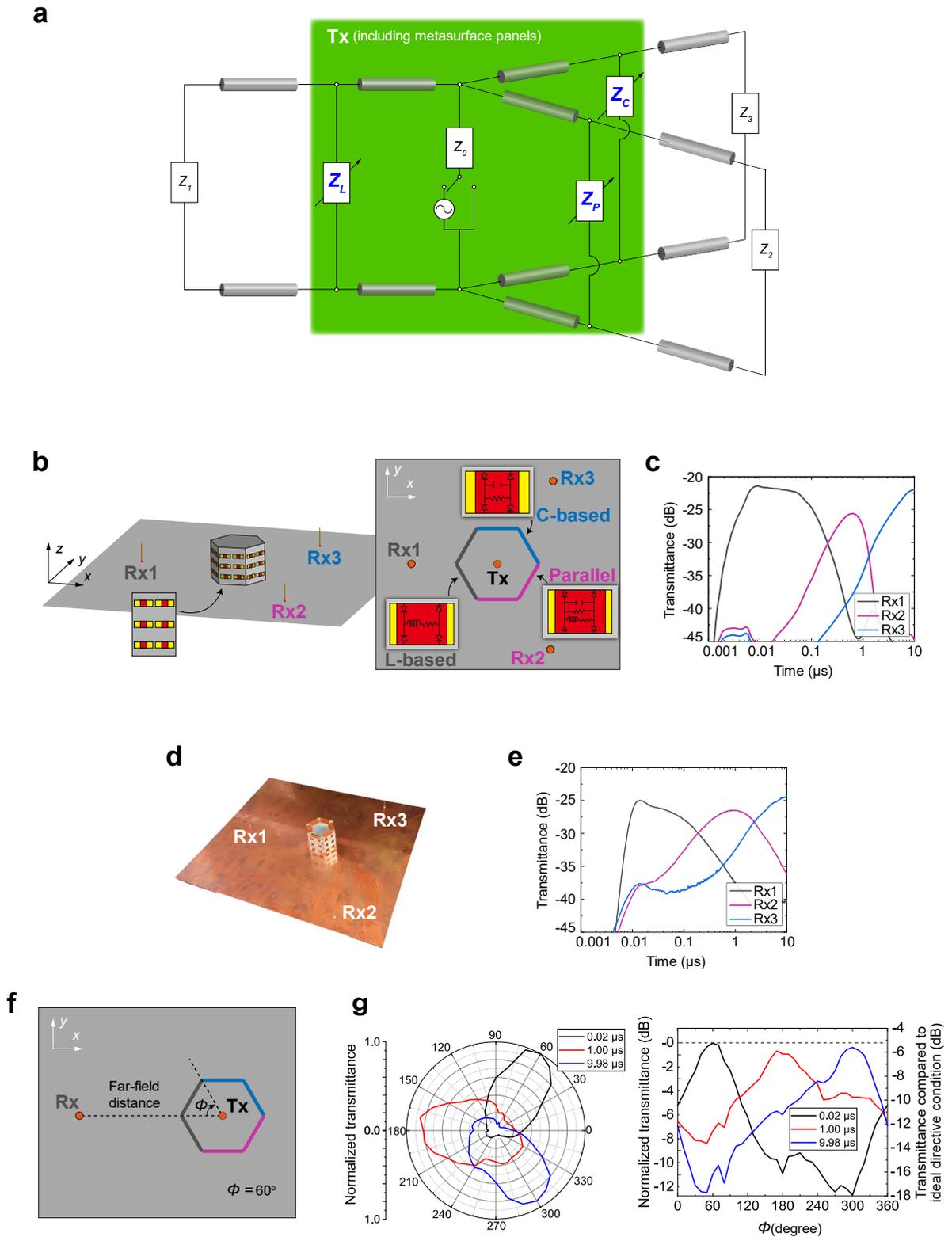

Fig. 3: System for free-space wave control. (a) Simplified equivalent circuit concept. A transmitter was expressed by an AC source, a switch and input impedance $Z_0$. C-based, L-based and parallel waveform-selective metasurfaces were respectively represented by variable shunt impedances $Z_C$, $Z_L$ and $Z_P$ between transmission lines that correspond to free space or vacuum. $Z_1$, $Z_2$ and $Z_3$ were impedances of three receivers. (b) Simulation model. A grounded

monopole transmitter (effectively working as a dipole suspended in free space) was surrounded by six panels, comprising two C-based, two L-based and two parallel-type waveform-selective transmitting metasurfaces (or slit structures introduced in Figure S13 and Figure S14). The distance between the transmitter (Tx) and the receivers (Rx1-Rx3) was 200 mm. (c) Simulated transmittance at 3.85 GHz with 30 dBm (see Figure S16 for frequency-domain profiles). (d) Experimental sample designed based on the simulation model above. (e) Experimental measurement results at 3.85 GHz with 30 dBm. (f) Schematic for measurement of the antenna radiation pattern where the metasurface hexagonal prism was rotated counter clock-wise forming an angle $\phi$ to the Tx-Rx line. In the current schematic $\phi = 60º$. Also, in this schematic, the C-based, L-based and parallel waveform-selective metasurface panels were represented by the same colours as the ones used in (b). (g) Time-varying radiation patterns of the metasurface antenna in (left) polar coordinate system and (right) Cartesian coordinate system, where the right axis represents comparison to ideal directive condition. Detailed information on the ideal directive condition is presented in Figure S17.

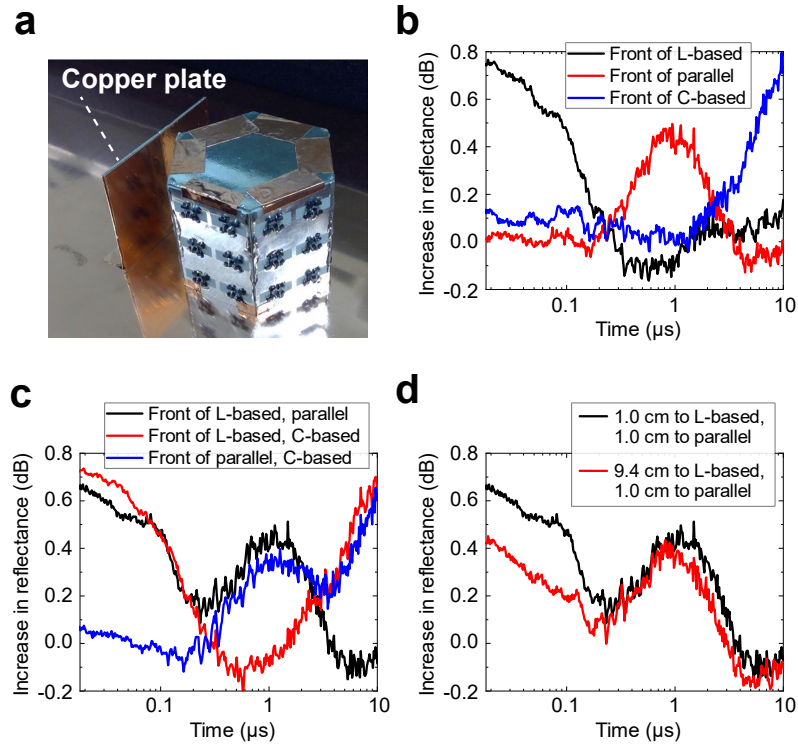

Fig. 4: Passive variable sensor to detect location of scattering objects. (a) Antenna system used in Fig. 3 in close proximity to a copper plate (70 mm wide and 51 mm tall). The distance between the copper plate and the closest waveform-selective metasurface panels was set to 1.0 cm as a default value. The frequency and power were 3.85 GHz and 30 dBm, respectively. (b) Measurement results when the copper plate was deployed in front of either the C-based, L-based or parallel waveform-selective metasurface panels. Additional measurement results using two copper plates with (c) the same distances and (d) different distances. In (d) the distance between one of the copper plates and the L-based waveform-selective metasurface was set to either 1.0 cm or 9.4 cm, while the distance between another plate and the parallel waveform-selective metasurface was fixed at 1.0 cm.

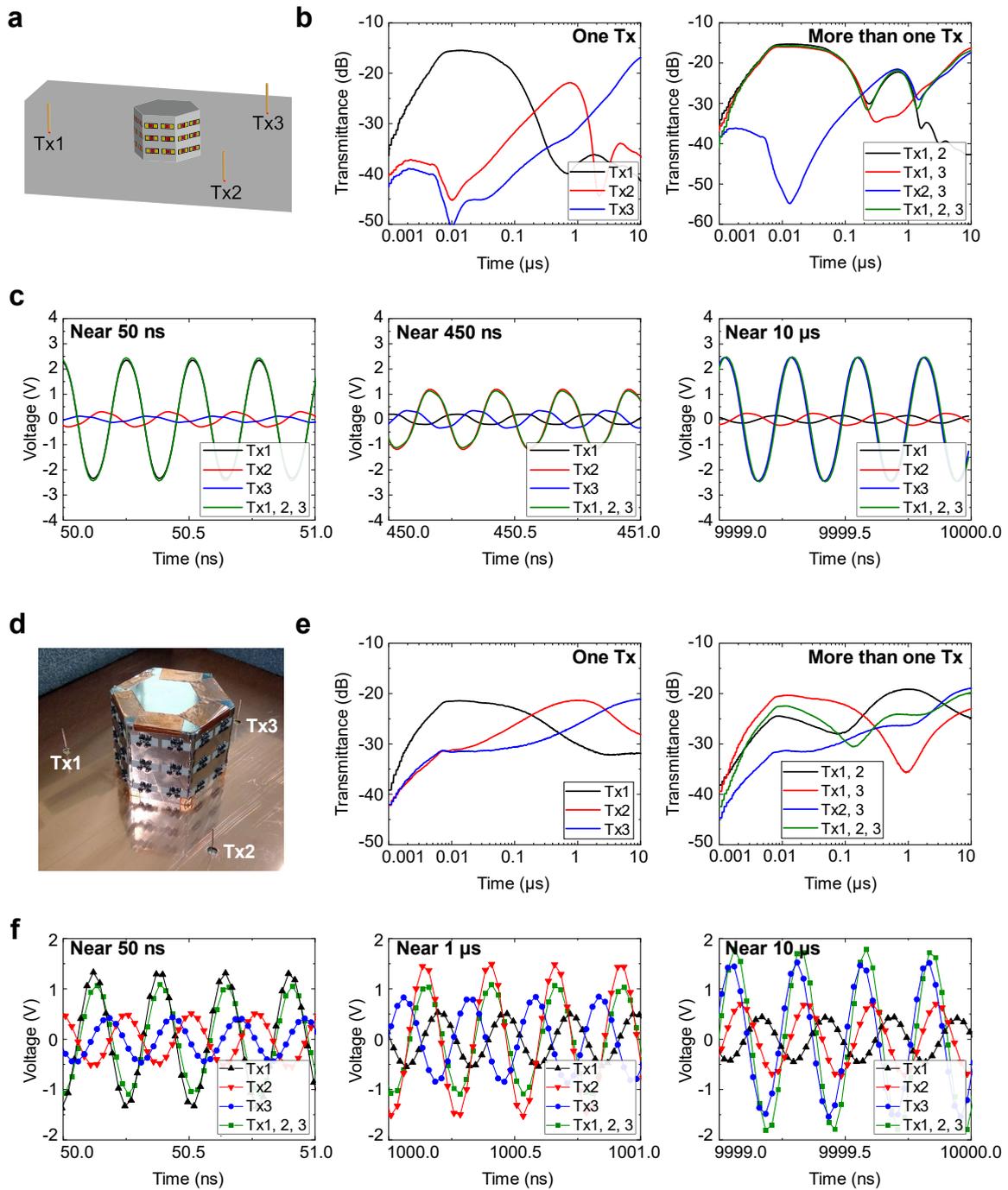

Fig. 5: Selective reception under simultaneous incidences. (a) Simulation model using three external monopoles as transmitters. The centre monopole selectively received different signals due to the presence of different waveform-selective transmitting metasurfaces. Frequency and input power were 3.85 GHz and 30 dBm, respectively. (b) Transmittances using (left) only one external transmitter and (right) more than one external transmitter. (c) Received voltages in different time slots. (d) Experimental sample and corresponding (e) transmittances and (f) received voltages with input power adjusted to 36 dBm. The distance between the transmitters (Tx1-Tx3) and the receiver (Rx) was 70 mm. Additional simulation results are shown in Figure S23, where the distance between the transmitters and the receiver was increased to 200 mm to represent a more realistic far-field distance.

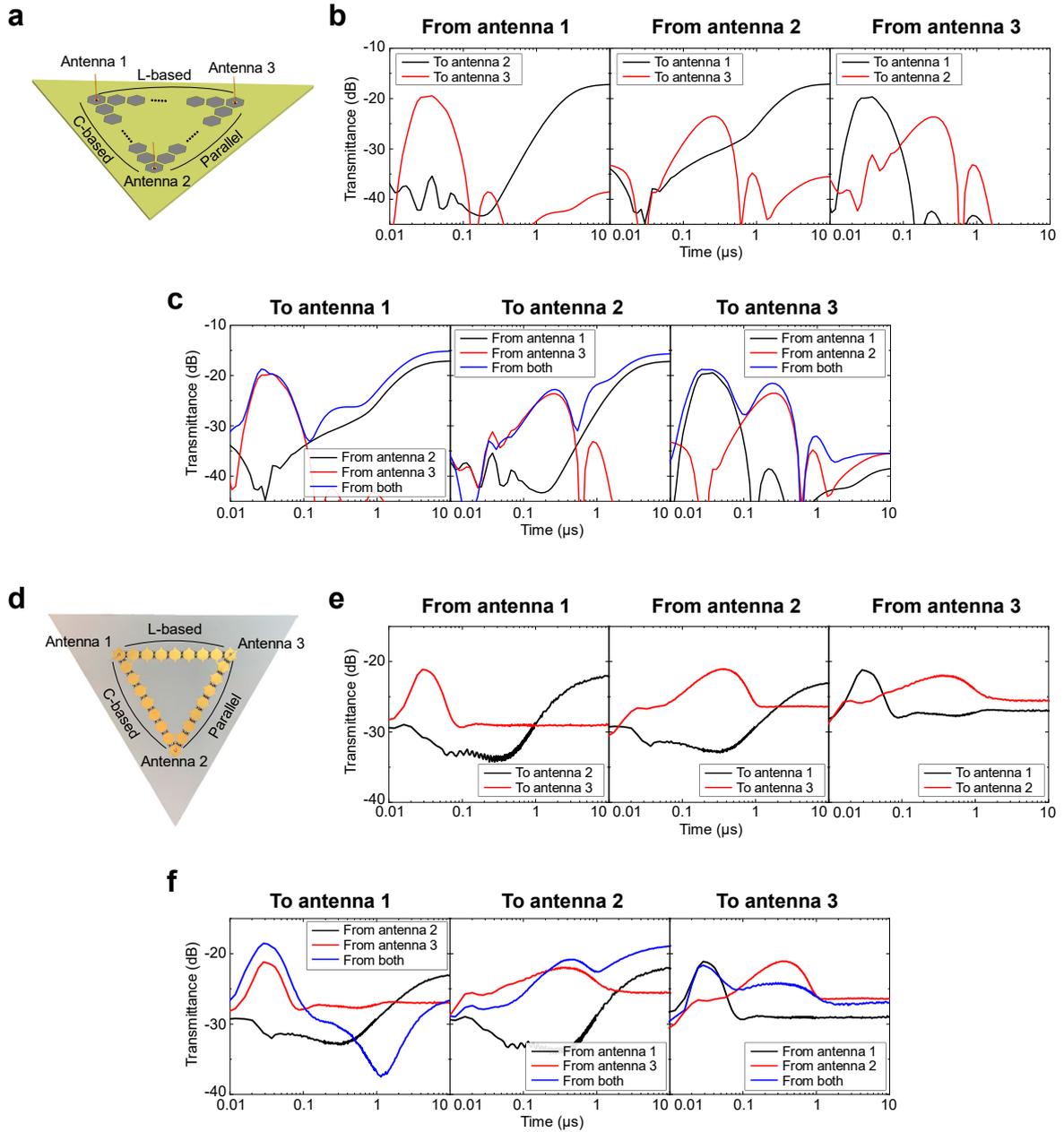

Fig. 6: Mutually pulse-width-selective communication system. (a) Simulation model using three antennas connected by either C-based, L-based or parallel waveform-selective metasurface line. The frequency and input power were set to 2.36 GHz and 10 dBm, respectively. (b) Simulation results of transmittances from (left) antenna 1, (centre) antenna 2 and (right) antenna 3. (c) Simulation results of transmittances to (left) antenna 1, (centre) antenna 2 and (right) antenna 3. (d) Measurement sample. The frequency and input power were set to 2.38 GHz and 15 dBm, respectively. (e) Measurement results of transmittances from (left) antenna 1, (centre) antenna 2 and (right) antenna 3. (f) Measurement results of transmittances to (left) antenna 1, (centre) antenna 2 and (right) antenna 3.

# Supplementary Information

**Pulse-Driven Self-Reconfigurable Meta-Antennas**

*Daiju Ushikoshi, Riku Higashiura, Kaito Tachi, Ashif Aminulloh Fathnan, Suhair Mahmood, Hiroki Takeshita, Haruki Homma, Muhammad Rizwan Akram, Stefano Vellucci, Jiyeon Lee, Alessandro Toscano, Filiberto Bilotti, Christos Christopoulos and Hiroki Wakatsuchi*

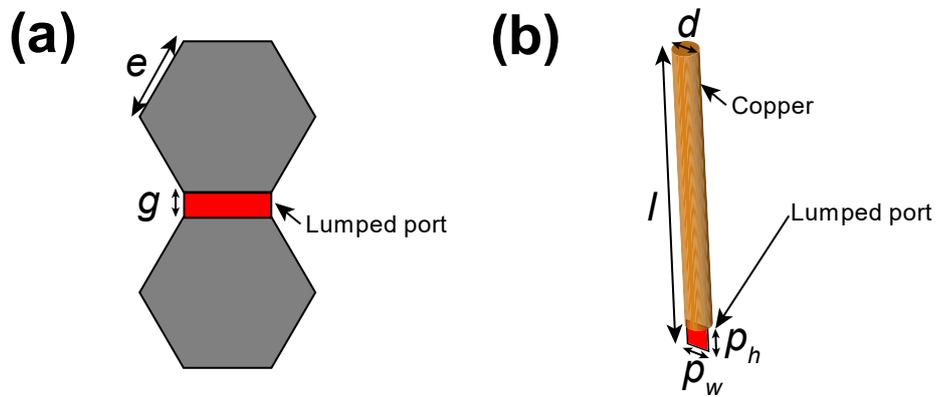

Figure S1: Design of the waveform-selective metasurface line and monopoles used in Fig. 1. Their design parameters are given in Table S1. Additionally, the circuit values of Fig. 1b are shown in Table S2. The substrate was Rogers3010 (1.27 mm thick).

Table S1: Design parameters of the waveform-selective metasurface model shown in Figure S1.

| Parameter | Length [mm] |
|---|---|
| $e$ | 12 |
| $g$ | 1 |
| $l$ | 30 |
| $d$ | 1 |
| $p_h$ | 1 |
| $p_w$ | 1 |

Table S2: Circuit parameters of the waveform-selective metasurface model used in Fig. 1a to Fig. 1f.

| Parameter | Value (self-resonant frequency) |
|---|---|
| $C$ | 1 nF (200 MHz) |
| $R_C$ | 100 kΩ |
| $L$ | 100 µH (10 MHz) |
| $R_L$ | 5.5 Ω |
| $C_P$ | 100 pH (750 MHz) |
| $R_{CP}$ | 100 kΩ |
| $L_P$ | 1 mH (2.4 MHz) |
| $R_{LP}$ | 24 Ω |

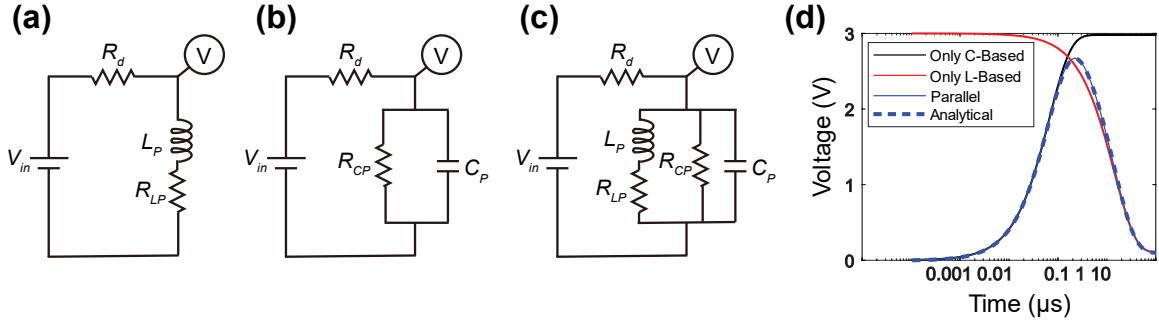

Figure S2: Schematics of DC circuit simulations where transient solver was used for (a) L-based, (b) C-based and (c) parallel type circuits. Here the equivalent diode resistance was represented by $R_d$ and set to $R_d = 680$ Ω. The input voltage $V_{in}$ was 3V. Transient circuit components were identical to those used in the parallel waveform-selective metasurface presented in Table S4, i.e., $L_P = 1$ mH, $R_{LP} = 24$ Ω, $C_P = 100$ pF, and $R_{CP} = 100$ kΩ. (d) Simulated voltages over time for the C-based, L-based and parallel circuits using the aforementioned circuit parameters. The dashed line is the analytical result obtained from Eq. 10 of Ref [49].

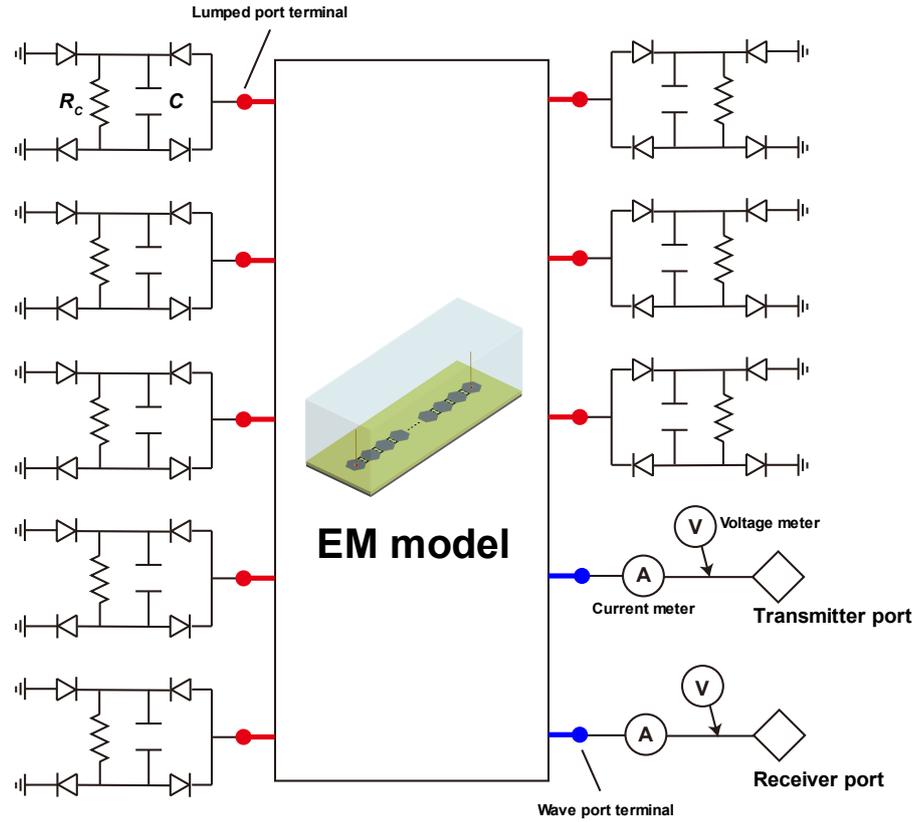

Figure S3: Schematic representing the co-simulation method in Fig. 1. An EM model was used in a circuit simulator and connected to lumped components and a signal source. Transmitter and receiver ports were used as the Tx and Rx. $R_C$s and $C$s were used for the C-based waveform-selective metasurface line, while other components were alternatively used for the L-based and parallel waveform-selective metasurface lines. More detail is provided under "Simulation Method" in the Methods Section.

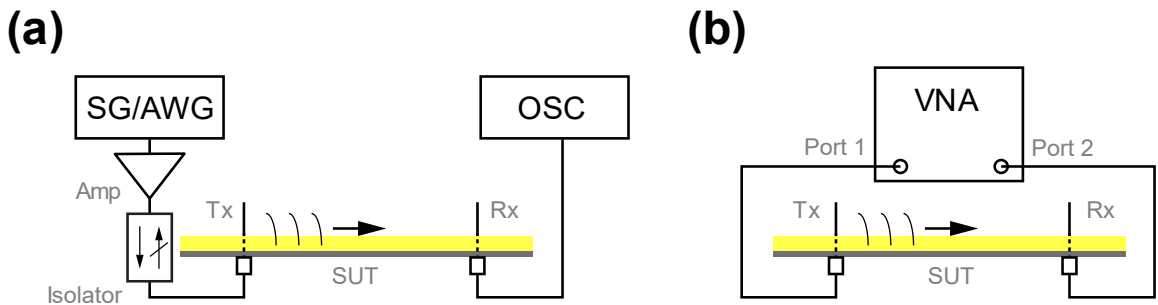

Figure S4: Setups for measuring (a) time-domain characteristics and (b) frequency-domain characteristics of Fig. 1 and Figure S8.

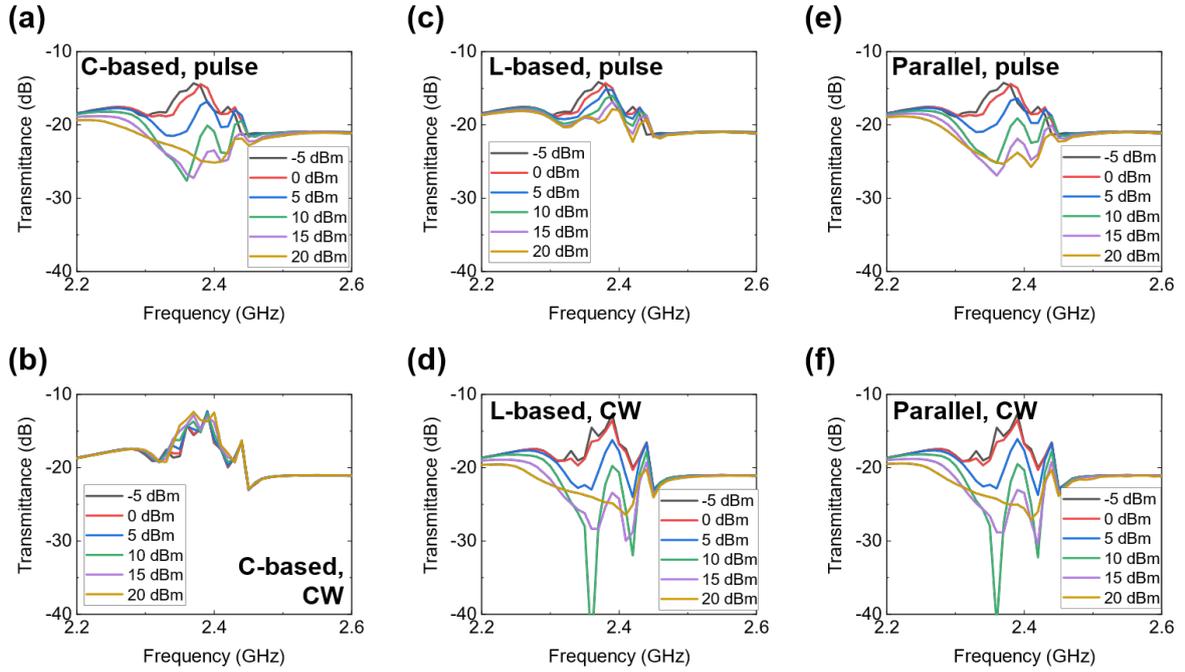

Figure S5: Simulation results of Fig. 1e with various power levels. Transmittances with (a, b) C-based, (c, d) L-based and (e, f) parallel waveform-selective metasurface lines. The top panels show the transmittances for 50-ns short pulses, while the bottom panels represent those for CWs. These results show that almost identical transmittances were obtained with low-power signals (e.g., -5 dBm) despite the differences in the input waveforms and the waveform-selective metasurfaces. This was because the power intensity was not large enough to turn on the diodes used. However, the transmittances differed from each other with sufficiently large power (e.g., 10 dBm).

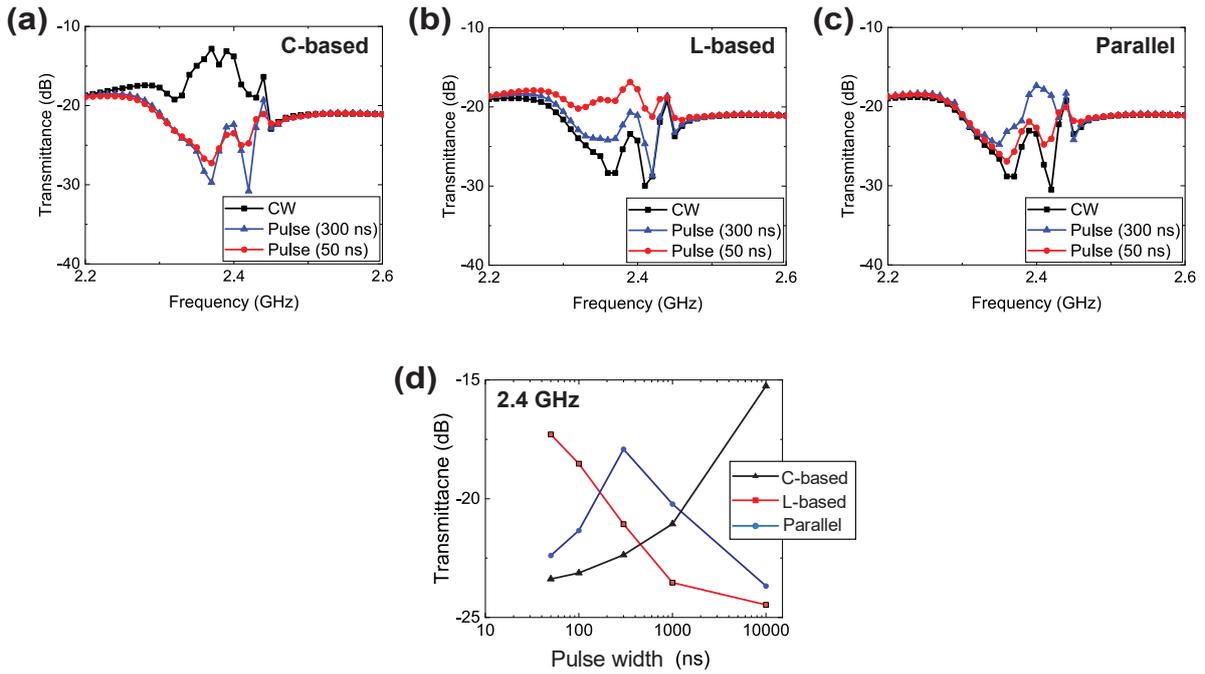

Figure S6: Simulation results of Fig. 1e with variation in pulse width. Comparison of transmittance over frequency using CWs, 50-ns pulses and 300-ns pulses for (a) C-based, (b) L-based and (c) parallel waveform-selective metasurfaces. (d) Transmittance with respect to pulse width variation (50 ns, 100 ns, 300 ns, 1 µs and 10 µs) at 2.4 GHz for C-based, L-based and parallel waveform-selective metasurfaces.

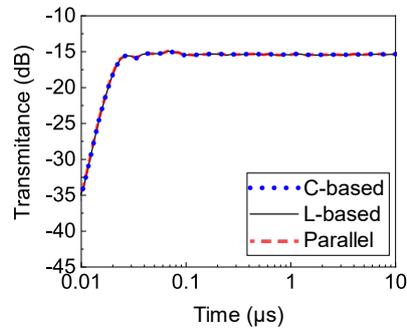

Figure S7: Simulation result of the left panel of Fig. 1f with -5 dBm. The signal generated from Tx took less than 100 ns to reach Rx for all three cases. Additionally, the three transmittances became constant and identical to each other.

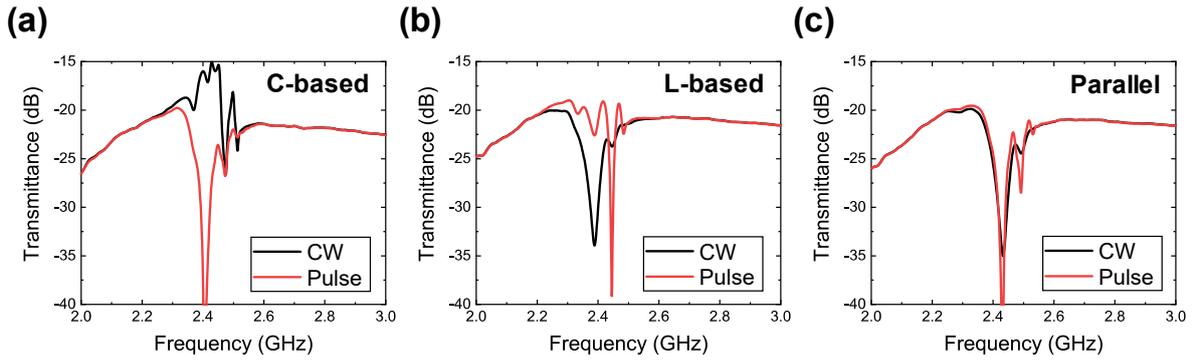

Figure S8: Measurement results of Fig. 1f in the frequency domain with 10 dBm. The results using (a) the C-based, (b) L-based and (c) parallel waveform-selective metasurface. In (a) transmittance for a CW was relatively larger than that for a 50-ns short pulse near 2.4 GHz due to the transient absorbing mechanism of the C-based waveform-selective metasurface. In contrast, the L-based waveform-selective metasurface more strongly absorbed a short pulse than a CW at the same frequency region, which appeared as a smaller transmittance for the CW in (b). Since both types of the circuit configurations were used in (c), the parallel waveform-selective metasurface reduced transmittances for a CW and a short pulse near 2.4 GHz.

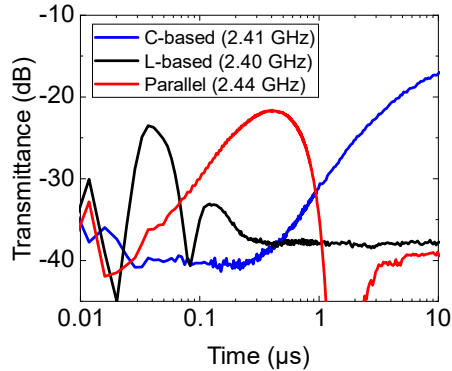

Figure S9: Transmittances of the measurement samples shown in the right panel of Fig. 1f with more optimized frequencies. The input power was fixed at 10 dBm. Results for other input powers and frequencies are seen in Figure S10 and Figure S11.

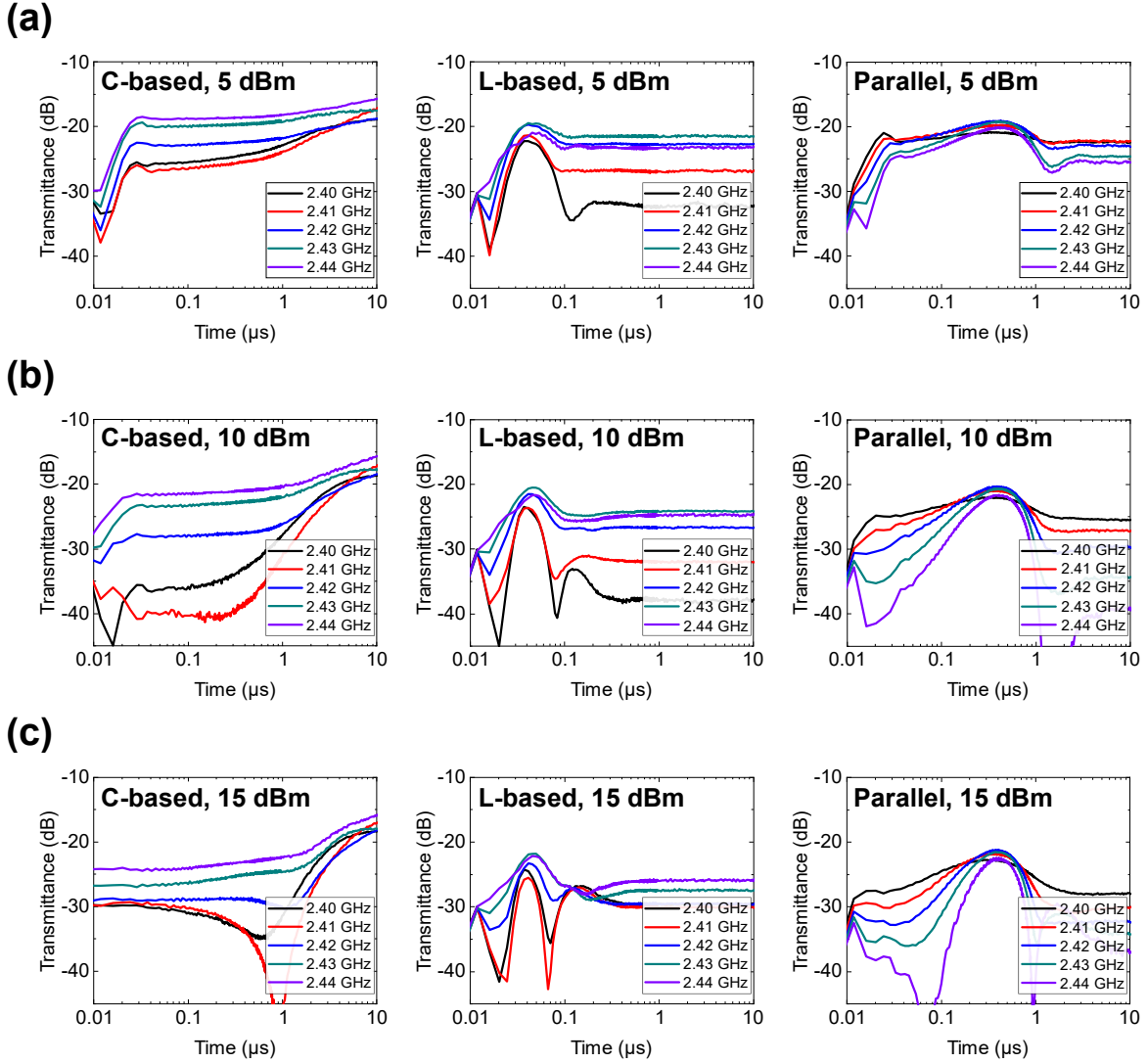

Figure S10: Transmittances of the measurement samples used in the right panel of Fig. 1f with different frequencies. The results using (a) 5 dBm, (b) 10 dBm and (c) 15 dBm. From the left panels to the right panels, the results correspond to the transmittances of the C-based, L-based and parallel waveform-selective metasurfaces, respectively. These results indicate that the transmittances tended to most effectively vary when the input power was set to 10 dBm or 15 dBm. At 2.40 GHz and 2.41 GHz, for instance, when the input power was 5 dBm, the transmittance of the C-based waveform-selective metasurface was approximately -25 dBm near 50 ns and 500 ns (see the black and the red curves of the left panel of (a)). However, by increasing the input power to 10 or 15 dBm, the transmittance became lower than -30 dBm. Note that near 50 ns and 500 ns, the L-based and the parallel waveform-selective metasurfaces increased the transmittances in the centre and the right panels of (a), which means that with 10 dBm and 15 dBm the difference between the three structures was maximized. In Figure S11, these results are plotted in accordance with frequency and input power.

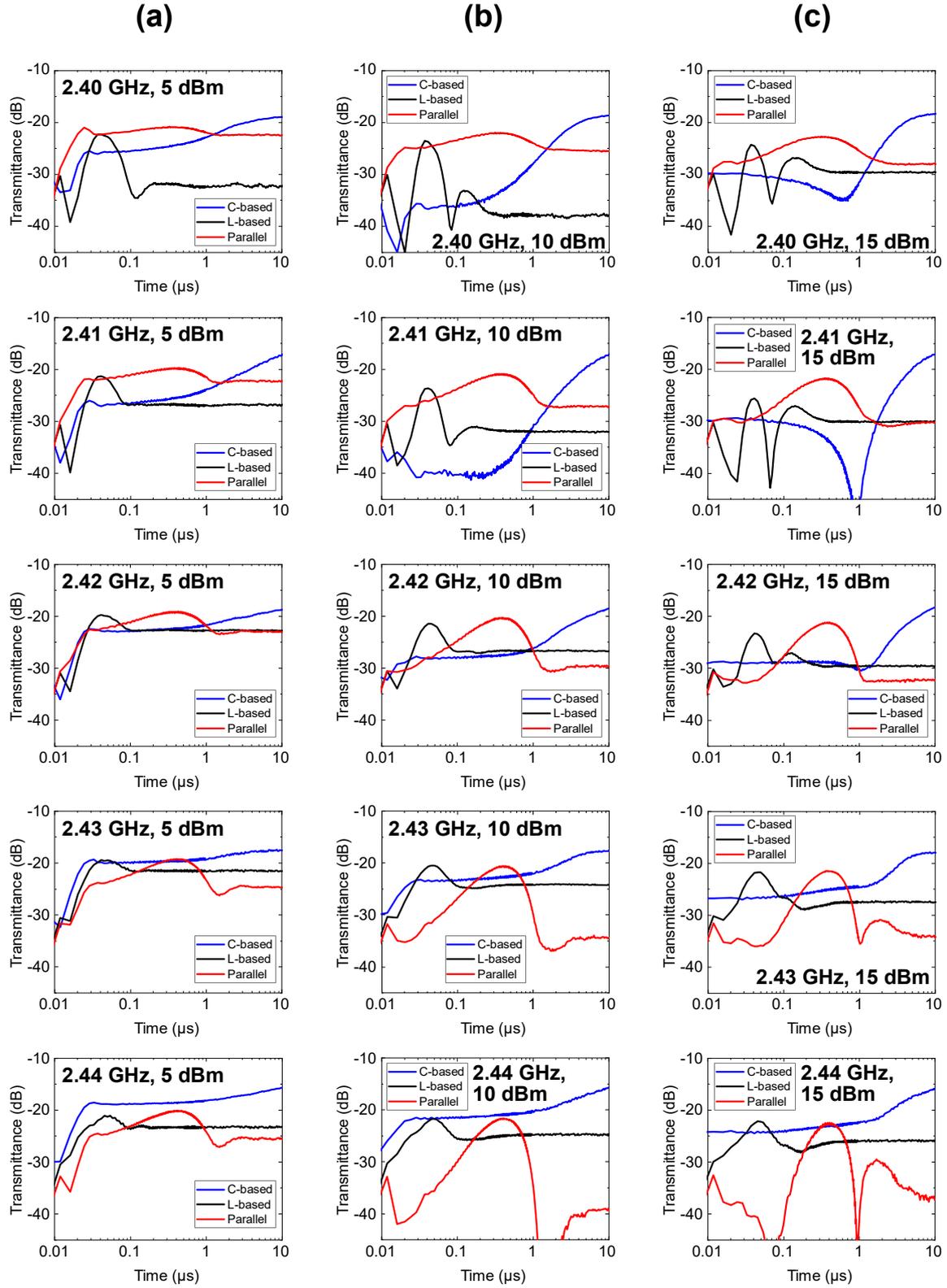

Figure S11: Transmittances of the measurement samples used in the right panel of Fig. 1f with different waveform-selective metasurfaces. The results using (a) 5 dBm, (b) 10 dBm and (c) 15 dBm. From the top panels to the bottom panels, the results correspond to the transmittances at 2.40, 2.41, 2.42, 2.43 and 2.44 GHz, respectively. In Figure S10, these results are plotted in accordance with input power and type of waveform-selective metasurface.

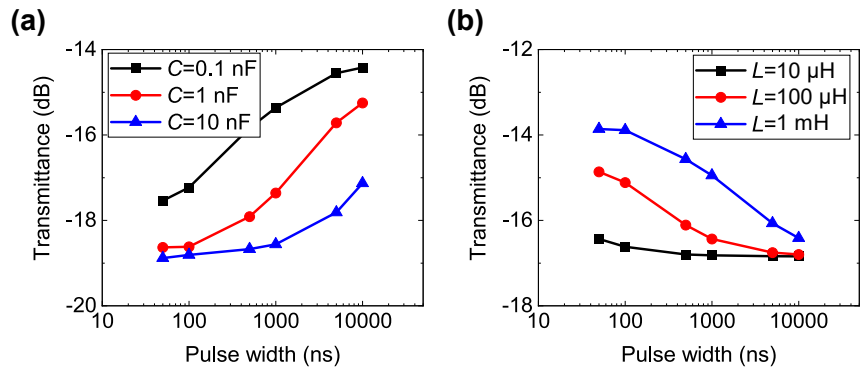

Figure S12: Simulated transmittances of (a) the C-based and (b) the L-based waveform-selective metasurface lines used in the left panel of Fig. 1f with various circuit values. By increasing *C* and *L*, the transient responses were shifted to a large time range.

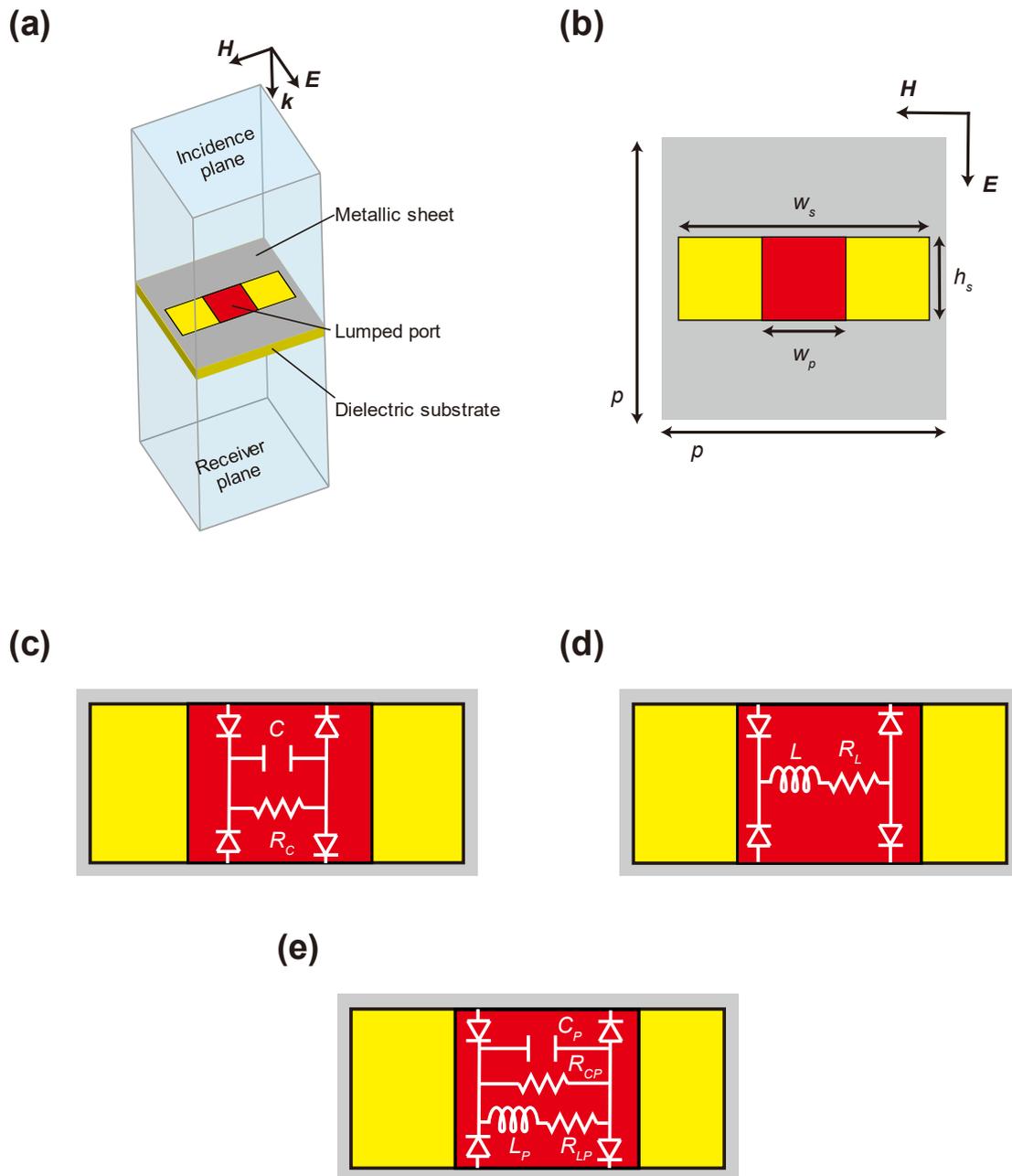

Figure S13: Single unit cell of the waveform-selective transmitting metasurface (slit structure) used for Fig. 3. (a) Periodic unit cell. Periodic boundaries were applied to the incident $E$ and $H$ planes. (b) The front surface of the periodic unit cell. (c, d and e) Circuit configurations of C-based, L-based and parallel waveform-selective transmitting metasurfaces. The design parameters are shown in Table S3 and Table S4. The substrate was Rogers3010 (1.27 mm thick).

Table S3: Design parameters of the waveform-selective metasurface model used in Figure S13b.

| Parameter | Length [mm] |
|---|---|
| $p$ | 17 |
| $w_s$ | 15 |
| $h_s$ | 5 |
| $w_p$ | 5 |

Table S4: Circuit parameters of the waveform-selective metasurface model used in Figure S13c to Figure S13e.

| Parameter | Value (self-resonant frequency) |
|---|---|
| $C$ | 1 nF (200 MHz) |
| $R_C$ | 10 kΩ |
| $L$ | 100 μH (10 MHz) |
| $R_L$ | 5.5 Ω |
| $C_P$ | 100 pF (750 MHz) |
| $R_{CP}$ | 100 kΩ |
| $L_P$ | 1 mH (2.4 MHz) |
| $R_{LP}$ | 24 Ω |

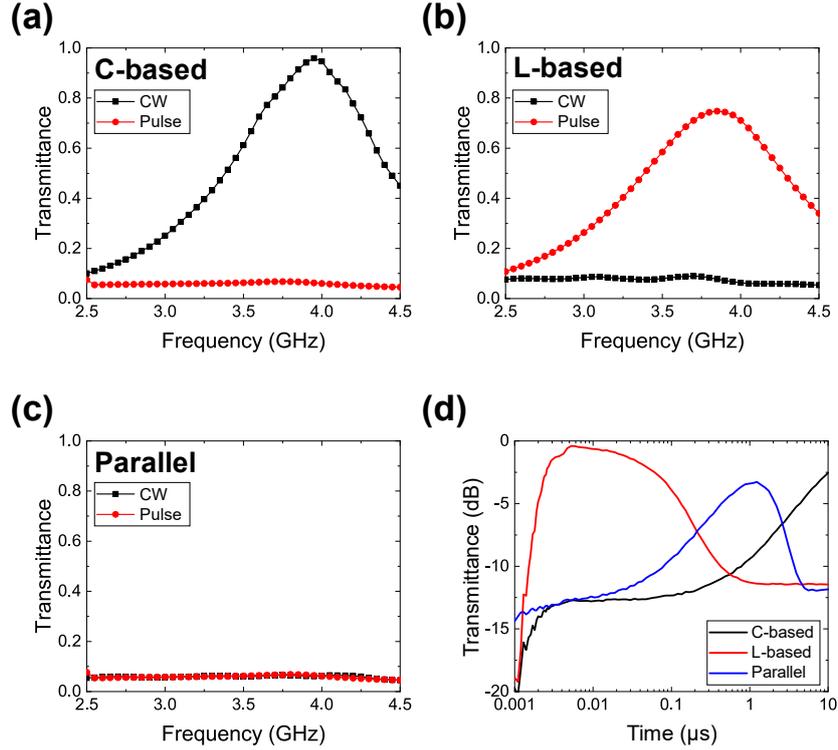

Figure S14: Simulated transmittances of the (a) C-based, (b) L-based and (c) parallel waveform-selective transmitting metasurfaces in Figure S13. The C-based waveform-selective metasurface more strongly transmitted CWs at approximately 4.0 GHz than short pulses, as the structure permitted induced electric charges to enter the diode bridge during an initial time period, which lowered the intensity of the intrinsic transmitting resonant mechanism of the slit structure. In contrast, the L-based waveform-selective metasurface showed more enhanced transmittance for short pulses than that for CWs. This was because at a steady state, the electromotive force of the inductor almost disappeared, which shortened the gap in the slit and weakened the intrinsic resonant mechanism of the structure. In the case of the parallel waveform-selective metasurface, both types of waveforms were poorly transmitted due to the presence of both types of circuit configurations. (d) Transient transmittances at 3.85 GHz. The incident power was set to 10 dBm. Due to the three different transient characteristics, the transmittances were maximized in different time slots.

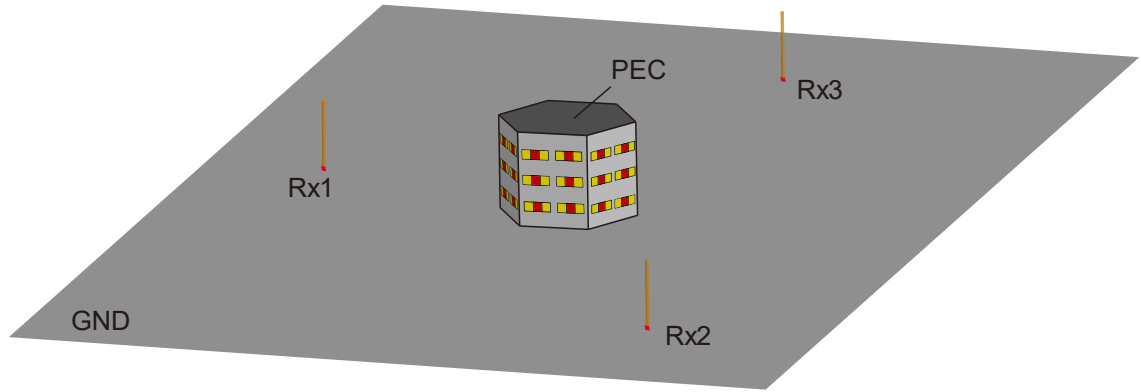

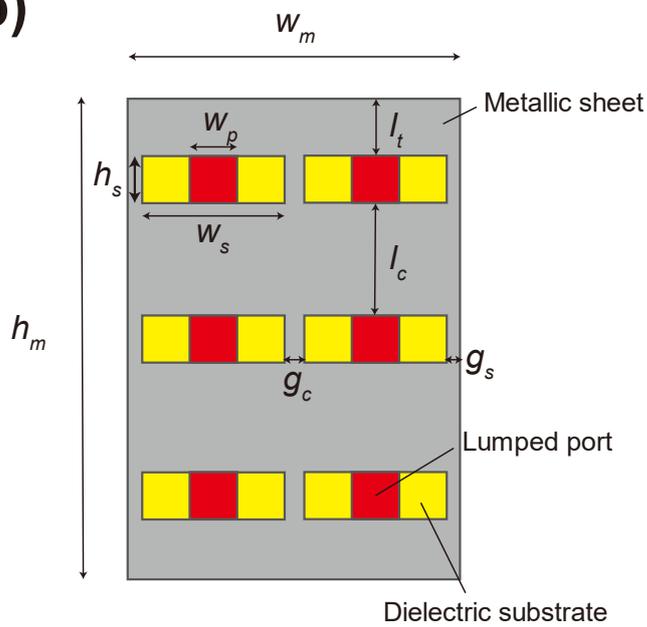
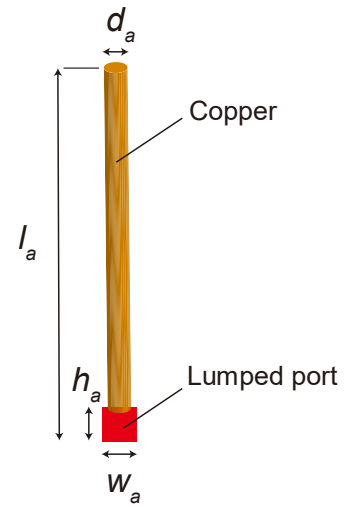

Figure S15: Antenna simulation model used for Fig. 3. Dimensions and circuit values are given in Table S5 and Table S6. The substrate was Rogers3010 (1.27 mm thick). Rx1, Rx2 and Rx3 were respectively placed in front of L-based, parallel and C-based waveform-selective metasurface panels.

Table S5: Design parameters of the waveform-selective metasurface model used in Figure S15 (i.e., in Fig. 3).

| Parameter | Length [mm] |
|---|---|
| $w_m$ | 35 |
| $h_m$ | 51 |
| $w_s$ | 15 |
| $h_s$ | 5 |
| $w_p$ | 5 |
| $l_t$ | 6 |
| $l_c$ | 12 |
| $g_c$ | 2 |
| $g_s$ | 1.5 |
| $l_a$ | 18 |
| $d_a$ | 1 |
| $w_a$ | 1 |
| $h_a$ | 1 |

Table S6: Circuit parameters of the waveform-selective metasurface model used in Figure S15 (i.e., in Fig. 3).

| Parameter | Value (self-resonant frequency) |
|---|---|
| $C$ | 10 nF (2 GHz) |
| $R_C$ | 100 kΩ |
| $L$ | 100 μH (10 MHz) |
| $R_L$ | 5.5 Ω |
| $C_P$ | 1 nF (200 MHz) |
| $R_{CP}$ | 100 kΩ |
| $L_P$ | 1 mH (2.4 MHz) |
| $R_{LP}$ | 24 Ω |

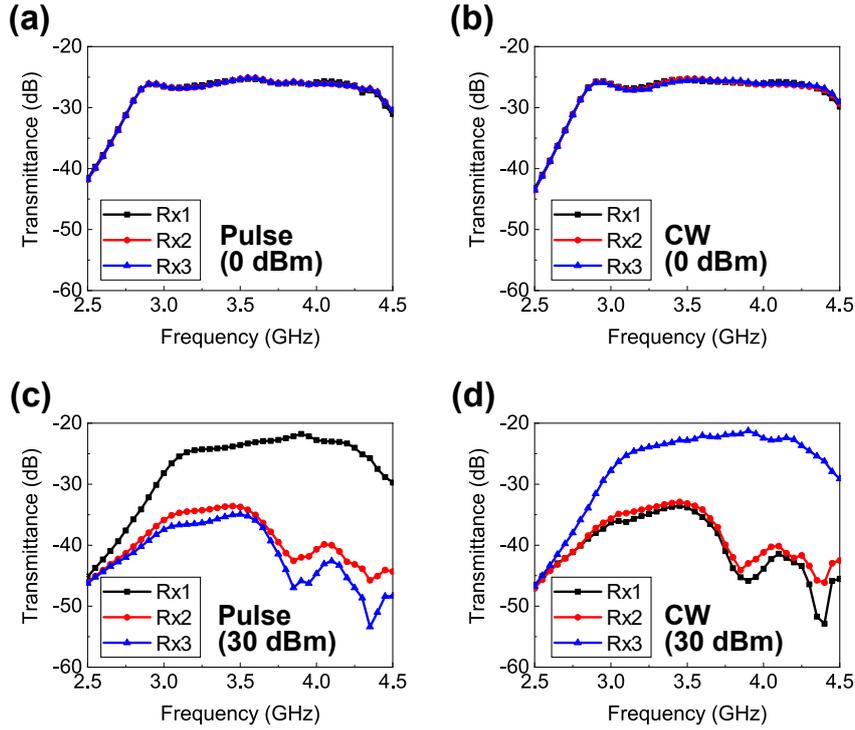

Figure S16: Simulated frequency dependence of the antenna shown in Fig. 3, composed of 3 different types of waveform-selective metasurfaces. (a) 0-dBm short pulses. (b) 0-dBm CWs. (c) 30-dBm short pulses. (d) 30-dBm CWs. As shown in (a) and (b), there was almost no difference between the transmittances for short pulses and those for CWs because the diodes used were not turned on. In contrast, (c) and (d) showed clear differences. Specifically, the transmittance of Rx1 was maintained at a large level for the 30-dBm short pulses between 3.2 and 4.5 GHz in (c), as the L-based waveform-selective metasurface deployed in front of Rx1 strongly transmitted the signal generated from the transmitter compared to the C-based and parallel waveform-selective metasurfaces. In the case of the 30 dBm CWs, the C-based waveform-selective metasurface effectively transmitted signals to Rx3, which appeared as a relatively large transmittance in (d). The parallel waveform-selective metasurface poorly transmitted both the 30-dBm pulses and the 30-dBm CWs due to the presence of both circuits, appearing as limited transmittances in (c) and (d).

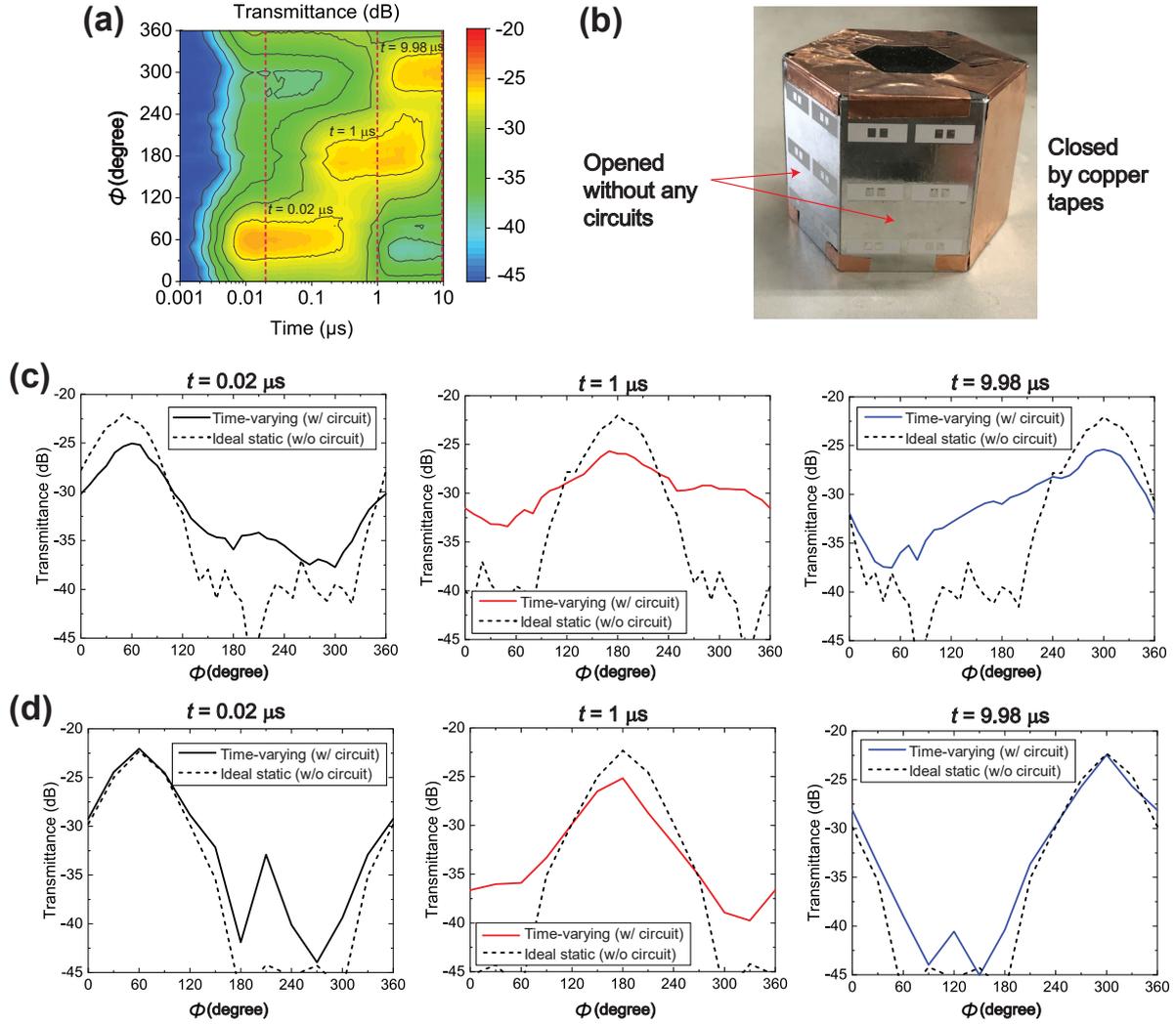

Figure S17. (a) Measured time-varying transmittance from the metasurface-based antenna shown in Fig. 3g with various far-field angles $\phi$. Three vertical dashed lines indicate time samples where the far-field patterns were presented in Fig. 3g (also presented here in Figure S17c). (b) Metasurface sample in ideal directive condition. Slits in two hexagonal sides were opened without circuits, while the others were fully closed by copper tapes. No time variation was observed in this ideal directive condition (static). (c) Comparison between the ideal and realized far-field patterns from the measurement results. In the right axis of the right panel of Fig. 3g, the measured far-field patterns were subtracted by the maximum level of the ideal directive condition, which gave a difference of around -5 dB. (d) Comparison between the ideal and realized far-field patterns from the simulation results. Here, a better agreement was obtained between the ideal and realized field power than those obtained in the measurement. This was presumably because additional losses introduced during the fabrication process reduced the measured transmittance.

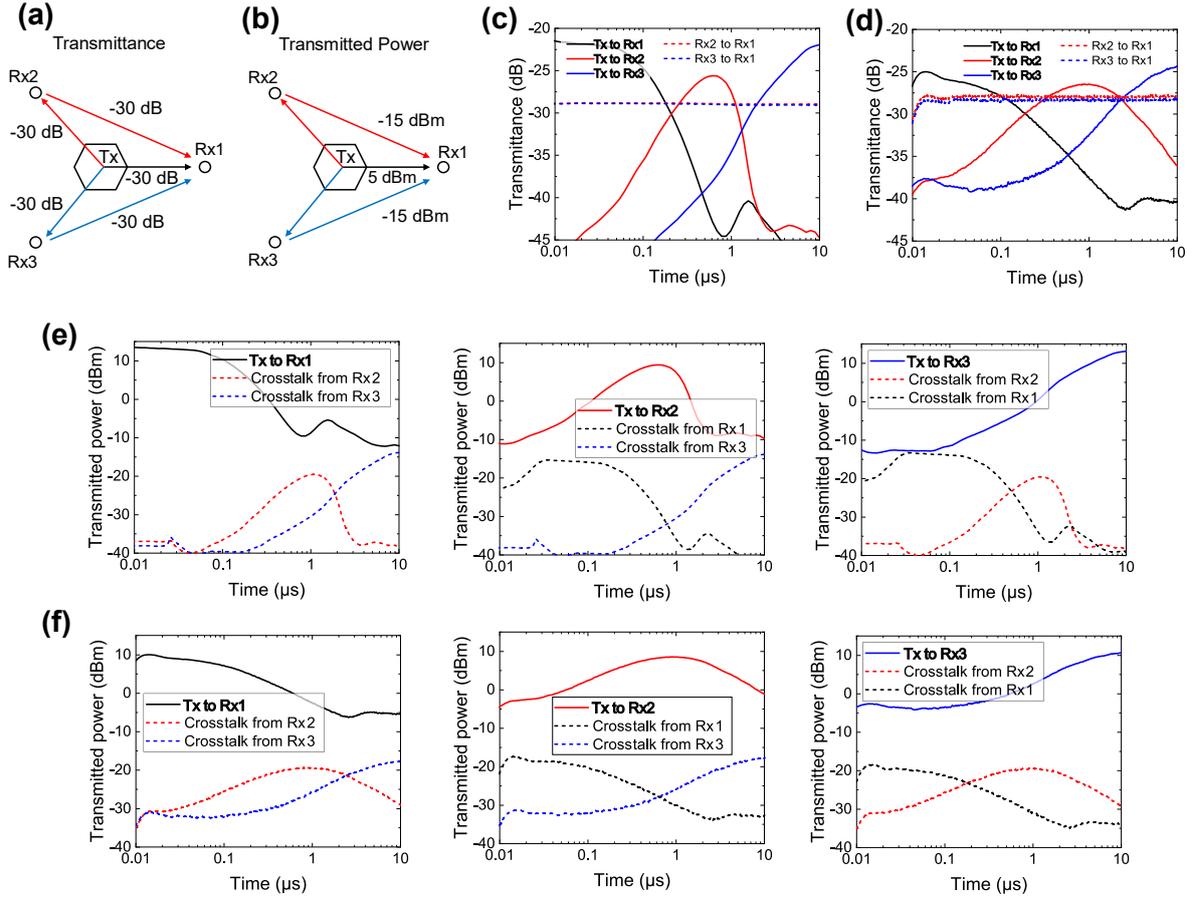

Figure S18: Characterization of crosstalk between external antennas with the proposed metasurface. (a) Schematic of interference paths. Rx1 also received secondary reflection from both Rx2 and Rx3. The estimated power in each path is indicated. (b) Transmitted power calculated using the transmittance in (a) when the source power was 35 dBm. While the main path from Tx to Rx1 gave 5 dBm power, the crosstalk from Rx2 and Rx3 was only -15 dBm. (c-d) Transmittance over time for all antennas outside the hexagonal metasurface prism in (c) simulation and (d) measurement. (e-f) Power received through the main path, i.e., corner directions of each metasurface panel, compared to the crosstalk both in (e) simulation and (f) measurement.

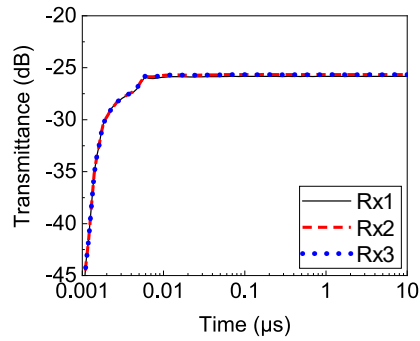

Figure S19: Simulated transient transmittance of the antenna shown in Fig. 3 and composed of 3 different types of waveform-selective metasurfaces at 3.85 GHz with 0 dBm. The simulation result shows that all the receivers observed the signal in almost the same manner as the diodes used were not turned on. This result also indicates that the signal took approximately 10 ns to reach the receivers. From 10 ns to 10 µs, the transmittance was fixed at a constant value since the transmitter behaved as an ordinary omnidirectional antenna. The related frequency-domain profiles are shown in Figure S16.

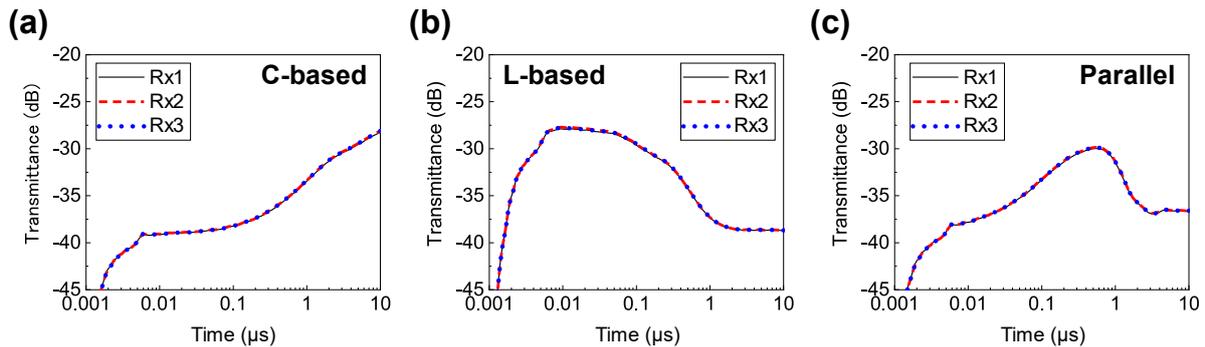

Figure S20: Simulated transient transmittance of the antenna shown in Fig. 3 but composed of only one of the three waveform-selective metasurfaces at 3.85 GHz with 30 dBm. The results using (a) C-based, (b) L-based and (c) parallel waveform-selective metasurfaces. Compared to Figure S18, although all the receivers received the signal in the same manner, the transmittance profile was shown to vary in the time domain.

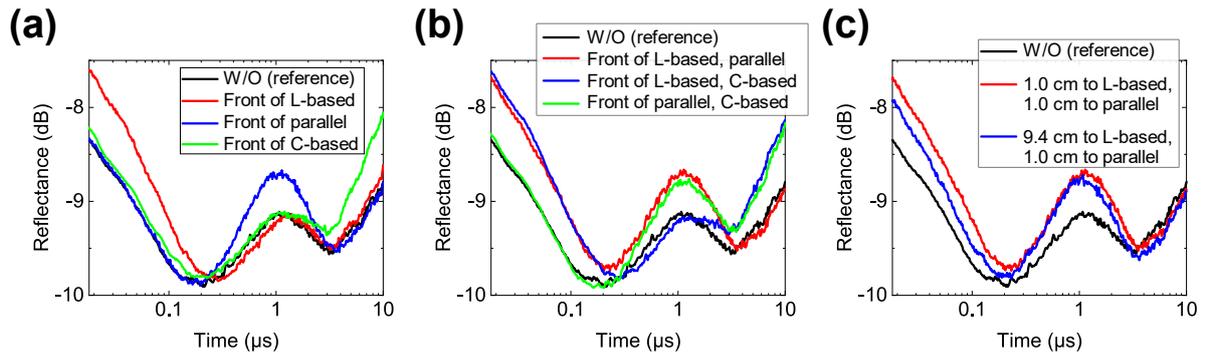

Figure S21: Measured reflectances of the passive variable sensor shown in Fig. 4. (a) Measurement results using only one copper plate. Measurement results using two copper plates with (b) the same distances and (c) different distances. Increases in reflectance of (a), (b) and (c) are seen in Fig. 4b, Fig. 4c and Fig. 4d, respectively.

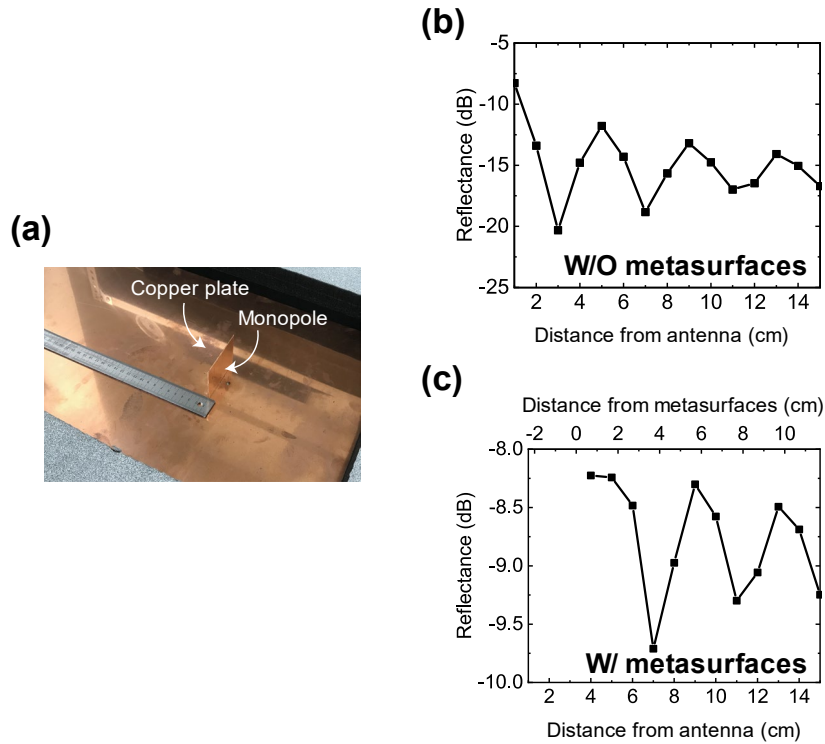

Figure S22: Measured reflectances of the passive variable sensor shown in Fig. 4 with various distances between the transmitting monopole and copper plate. (a) Image of the measurement without waveform-selective metasurface panels. Results (b) without and (c) with waveform-selective metasurface panels. In (c) the copper plate was positioned in front of the L-based waveform-selective metasurface panels. The result of (c) was obtained around 50 ns. These results indicate that there appeared a standing wave between the monopole and the copper plate, which affects the sensing performance. However, this influence can be mitigated by improving the impedance matching of the monopole. Nonetheless, these results show that the envelop curves in (b) and (c) gradually decreased by increasing the distance from the monopole antenna. Thus, if the impedance matching issue is addressed then the sensing performance is expected to be more enhanced to detect the distance from a scattering object.

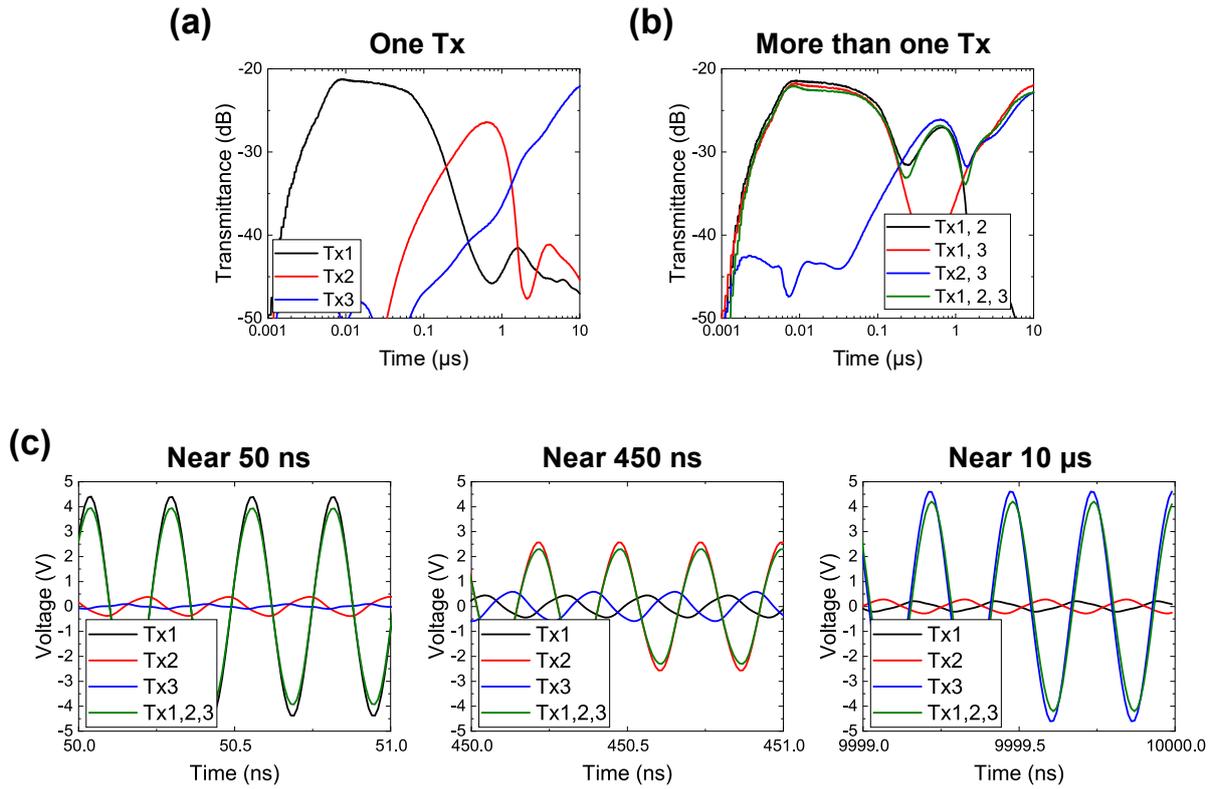

Figure S23: Additional simulation results for Fig. 5 with the distance between the transmitters and the receiver increased to 200 mm (representing a more realistic far-field distance). The input power was increased to 45 dBm to turn on the diodes used. (a) Transmittances with a single source. (b) Transmittances with more than one source. (c) Voltages in the time domain. Almost the same conclusion was drawn as that of Fig. 5 except minor discrepancies including different magnitudes in transmittances. More importantly, however, this figure also establishes that waveform-selective reception is achievable even under simultaneous incidences and with far-field radiation.

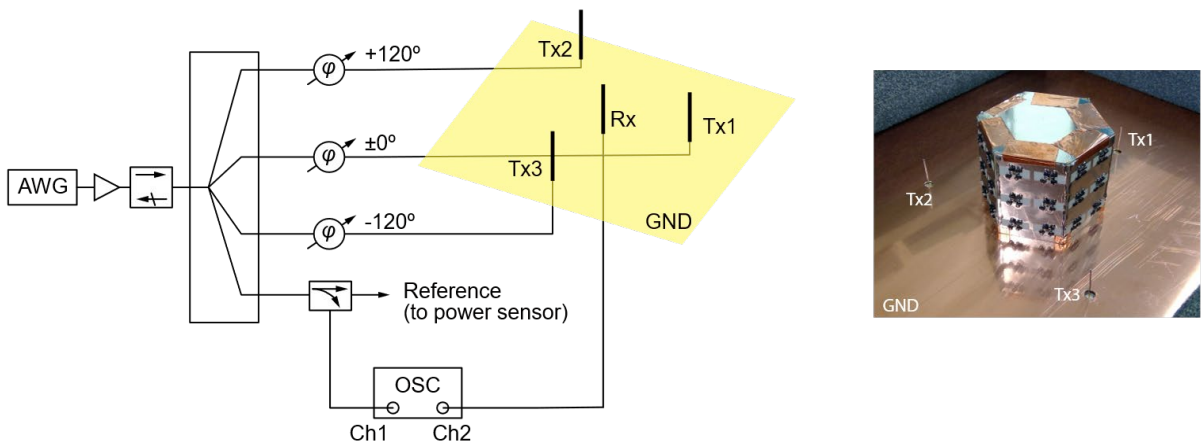

Figure S24: Measurement setup for the simultaneous incidences in Fig. 5. Tx1, Tx2 and Tx3 were respectively placed in front of L-based, parallel and C-based waveform-selective metasurface panels. A detailed description of the setup is presented under "Measurement Method" in the Methods Section.

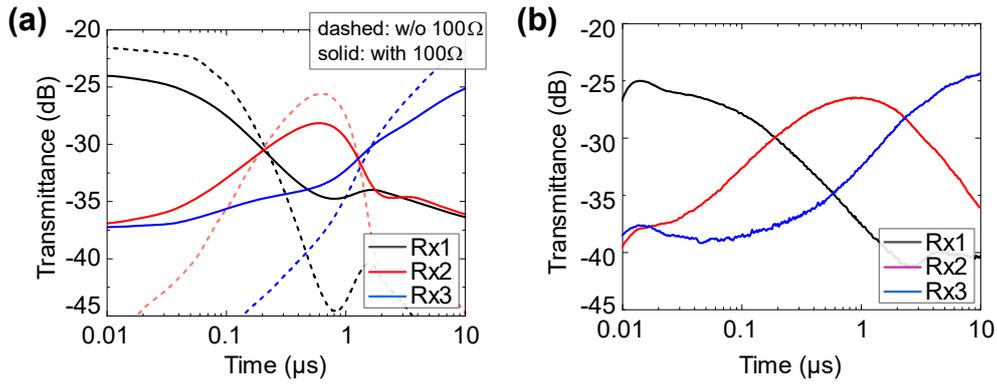

Figure S25. (a) Simulated transmittance of the metasurface antenna used in Fig. 3. The input frequency was set to 3.85 GHz with a 30-dBm power level. An additional resistor of 100 Ω was connected to each diode bridge in series (solid), compared to the original simulation result without the additional resistance (dashed). The resistance was deliberately added to resemble additional losses introduced during the fabrication process. (b) Measured transmittance using the same setup (also presented in Fig. 3c).

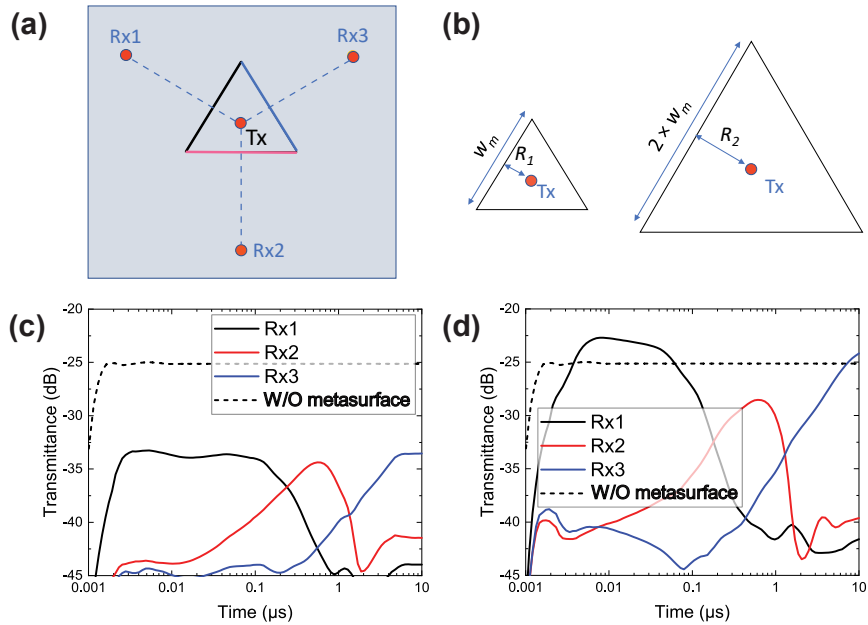

Figure S26: (a) Schematic of the equilateral triangle metasurface antenna simulation. (b) Comparison of two triangle metasurfaces having different inscribed radius of $R_1$ = 10.1 mm (~ 0.13$\lambda$) and $R_2$ = 20.1 mm (~ 0.26$\lambda$). In both cases the distance between the transmitter (Tx) and the receivers (Rx1-Rx3) was 200 mm. Transmittance of the metasurface antenna over time for the metasurface with (c) the small triangle shape and (d) the large triangle shape. The transmittance was significantly reduced in the small triangle case compared to the large triangle case due to the effect of coupling as well as standing wave that were amplified by the smaller distance between the monopole antenna and the metasurface panels (equivalent to the inscribed radius $R$).